\documentclass[review]{elsarticle}

\usepackage[english]{babel} 
\usepackage[T1]{fontenc}
\usepackage{lmodern} 
\usepackage[ansinew]{inputenc}
\usepackage{amsmath}
\usepackage{amssymb}
\usepackage{xfrac}
\usepackage{tikz}
\usepackage{tikz-dimline}
\usepackage{lineno}

\pgfplotsset{
compat=1.5,
legend image code/.code={
\draw[mark repeat=2,mark phase=2]
plot coordinates {
(0cm,0cm)
(0.15cm,0cm)        
(0.3cm,0cm)         
};
}
}
\usepackage{geometry}
\usepackage{blindtext}
\usepackage{float}
\usepackage{graphicx}
\usepackage{caption}
\usepackage{hyperref}
\usepackage{pgfplots}
\usepackage{array}
\usepackage{siunitx}
\usepackage{multirow}
\usepackage{adjustbox} 
\usepackage{subcaption}
\usepackage{bm}
\usepgfplotslibrary{groupplots,dateplot}
\usetikzlibrary{patterns,shapes.arrows,calc,external,decorations,shapes,positioning}
\tikzset{>=latex}
\pgfplotsset{compat=newest}

\begin{document}

\title{Consistent and symmetry preserving data-driven interface reconstruction for the level-set method}

\author{Aaron B. Buhendwa\corref{cor}}
\ead{aaron.buhendwa@tum.de}
\cortext[cor]{Corresponding author}

\author{Deniz A. Bezgin\corref{}}
\ead{deniz.bezgin@tum.de}

\author{Nikolaus A. Adams}
\ead{nikolaus.adams@tum.de}

\address{Department of Mechanical Engineering, Chair of Aerodynamics and Fluid Mechanics, Technical University of Munich, Germany}

\begin{frontmatter}
    \begin{abstract}
Recently, machine learning has been used to substitute parts of conventional computational fluid dynamics, e.g. the cell-face reconstruction in finite-volume solvers or the curvature
computation in the Volume-of-Fluid (VOF) method. The latter showed improvements in terms of accuracy for coarsely resolved interfaces, however at the expense of convergence and symmetry.
In this work, a combined approach is proposed, adressing the aforementioned shortcomings. 
We focus on interface reconstruction (IR) in the level-set method, i.e. the computation of the volume fraction and apertures. 
The combined model consists of a classification neural network,
that chooses between the conventional (linear) IR and the neural network IR depending on the local interface resolution. 
The proposed approach improves accuracy for coarsely resolved interfaces and recovers the conventional IR for high resolutions, yielding first order overall convergence. 
Symmetry is preserved by mirroring and rotating the input level-set grid
and subsequently averaging the predictions. 
The combined model is implemented into a CFD solver and demonstrated for two-phase flows. 
Furthermore, we provide details of floating point symmetric implementation and computational efficiency.
\end{abstract}
\end{frontmatter}
\journal{}
\section{Introduction}\label{introduction}

Multiphase flows with phase interfaces are omnipresent in nature and industrial applications. 
Therefore, the accurate simulation of such flows is subject of intense research.
Over the years, numerous methods for the simulation of such flows have been developed, most notably the Volume-of-Fluid (VOF) \cite{Hirt1981} and level-set \cite{Sussman1994b} method. 
Numerical models for surface tension in particular play an important role in the simulation of a wide range of multiphase flows. 
Here, the accurate computation of curvature is crucial.
In the VOF method, the interface is tracked by the volume fraction field, for which discrete schemes are not suitable for direct evaluation of the curvature.
In practice, the curvature is computed using a smoothly-varying function derived from a mapping of the volume fraction field, such as height functions \cite{Cummins2005}.

Recently, machine learning methods have been used to substitute certain parts of conventional computational fluid dynamics solvers.
Examples are cell-face reconstruction in finite-volume solvers \cite{Bezgin2021} and the curvature computation in VOF \cite{Qi2019, Patel2019a} and Smoothed-particle hydrodynamic \cite{Liu2021b} codes.
The general procedure can be described as follows: ML models are trained separately from the solver in a supervised fashion
on a dataset generated by generic, circular interface configurations. 
Subsequently, trained models are used within a CFD solver substituting the conventional curvature computation.
While the ML models display improvements with regard to accuracy for coarsely resolved interfaces, both symmetry and convergence are not provided.

In the level-set method, the interface is tracked by a scalar field whose values represent the signed distance to the interface. 
Since this is a smooth function, discrete schemes can easily be constructed to compute the curvature from the level-set field \cite{Popinet2018}. 
These schemes are fast and accurate (see \ref{appendix_curvature}). 
Even more so, if the curvature values computed are corrected for the distance of the interface from the cell center as described by Luo et al. \cite{Luo2015b}.
However, the interface reconstruction in the level-set method, i.e. the computation of the volume fraction and apertures, poses a bigger challenge.
Conventionally, the interface reconstruction in the level-set method is done via linear interpolation \cite{Hu2006} of the level-set field.
While this approach exhibits first order convergence and approaches (recovers) the exact solution for highly resolved (linear) interfaces, it yields poor results for curved interfaces that are coarsely resolved. 

In the present work, we introduce a consistent and symmetry-preserving data-driven interface reconstruction for the level-set method. 
We substitute the linear interface reconstruction with a regression neural network, improving the accuracy for coarsely resolved interfaces. Additionally, a classification neural network is 
trained to choose between neural network and linear interface reconstruction depending on the local resolution of the interface.
Thus, the combined model recovers the linear interface reconstruction for highly resolved interfaces and first order convergence.
Furthermore, symmetry is preserved by mirroring and rotating the neural network input with regard to all symmetries and subsequently averaging over the predictions.
The combined model is implemented into ALPACA \cite{Hoppe,Kaiser2019a}, a block-based multiresolution compressible two-phase solver, and validated on a rising bubble test case.
We provide implementation details with regard to floating point symmetry, speed and flexibility.

The remainder of the paper is structured as follows. Section \ref{methodology} presents a description of the linear and neural network interface reconstruction.
The data generation process and various data sets used to train and test the models are described in section \ref{data generation}. We display the accuracy of various model complexities outside the solver in section \ref{model perfomance},
followed by the performance inside the solver in section \ref{model perfomance solver}. The paper concludes with final remarks in section \ref{conclusion}.

\section{Methodology} \label{methodology}
\begin{figure}[t]
    \centering
    \begin{tikzpicture}

    \coordinate (NULL) at (0,0);
    \coordinate (B) at (5,5);
    \coordinate (C) at ($0.5*(B)$);
    \coordinate (D1) at ($0.125*(B)$);
    \coordinate (D2) at ($0.71*(B)$);

    \coordinate (D11) at ($0.1*(B)$);

    \coordinate (L) at ($1.1*(B|-NULL)+0.96*(B-|NULL)$);

    \filldraw[fill=black!10!white, line width=0.1pt] (NULL) -- (B|-NULL) -- ($(B|-NULL) + 0.16*(B-|NULL)$) -- ($(B|-NULL) + 0.16*(B-|NULL)$) arc (33.5:56.4:15) -- (B-|NULL) -- (NULL);

    \draw[line width=1pt] (0,0) rectangle (B);
    \draw[line width=0.5pt] (C |- NULL) -- (C |- B);
    \draw[line width=0.5pt] (C -| NULL) -- (C -| B);
    \draw[fill=red, draw=red] (C) circle (0.1);

    \draw[dashed] ($(C)+(C-|NULL)$) -- ($(C)+(B-|NULL)$);
    \draw[dashed] ($(C)+(C|-NULL)$) -- ($(C)+(B|-NULL)$);
    \draw[dashed] ($(C)-(C-|NULL)$) -- ($(C)-(B-|NULL)$);
    \draw[dashed] ($(C)-(C|-NULL)$) -- ($(C)-(B|-NULL)$);
    \draw[dashed] ($3*(C|-NULL)-(NULL|-C)$) -- ($-1*(C)$) -- ($3*(C-|NULL)-(NULL-|C)$) -- ($3*(C)$) -- cycle;

    \node[left, xshift=-0.8cm] at (D2 -| C) {$\color{blue}\Delta\Gamma_{i,j}$};

    \draw[fill=red, draw=red] (C) circle (0.1) node[below left] {$(i, j)$};
    \draw[fill=red, draw=red] ($(C)+(B-|NULL)$) circle (0.1) node[below right] {$(i, j+1)$};
    \draw[fill=red, draw=red] ($(C)+(B|-NULL)$) circle (0.1) node[below right] {$(i+1, j)$};
    \draw[fill=red, draw=red] ($(C)-(B|-NULL)$) circle (0.1) node[below left] {$(i-1, j)$};
    \draw[fill=red, draw=red] ($(C)-(B-|NULL)$) circle (0.1) node[below right] {$(i, j-1)$};
    \draw[fill=red, draw=red] ($-1*(C)$) circle (0.1) node[below left] {$(i-1, j-1)$};
    \draw[fill=red, draw=red] ($3*(C)$) circle (0.1) node[below right] {$(i+1, j+1)$};
    \draw[fill=red, draw=red] ($3*(C-|NULL)-(NULL-|C)$) circle (0.1) node[below left] {$(i-1, j-1)$};
    \draw[fill=red, draw=red] ($3*(C-|NULL)-(NULL-|C)$) circle (0.1) node[below left] {$(i-1, j+1)$};
    \draw[fill=red, draw=red] ($3*(C|-NULL)-(NULL|-C)$) circle (0.1) node[below right] {$(i+1, j-1)$};

    \draw[red, line width=1pt] ($1.1*(B|-NULL)$) arc (30:60:15);

    \draw[blue, line width=1pt] (B -| D11) -- (D11 -| B);

    \draw[blue, line width=1pt, dash dot] (B -| D1) -- (D2 -| C);
    \draw[blue, line width=1pt, dash dot] (D2 -| C) -- (C -| D2);
    \draw[blue, line width=1pt, dash dot] (D1 -| B) -- (C -| D2);

    \coordinate (DIST) at (0.5,0.5);
    \dimline[extension start length=1cm, extension end length=1cm,extension style={black}, label style={above=0.5ex}] {(-0.5,0)}{($(NULL|-B) - (0.5,0)$)}{$A_{i-\frac{1}{2},j}=1.0$};
    \dimline[extension start length=-1cm, extension end length=-1cm,extension style={black}, label style={below=0.5ex}] {(0.0,-0.5)}{($(NULL-|B) - (0.0,0.5)$)}{$A_{i,j-\frac{1}{2}}=1.0$};
    \dimline[extension start length=0.5cm, extension end length=0.5cm,extension style={black}, label style={above=0.5ex}] {($(NULL|-B) + (DIST -| NULL)$)}{($(NULL|-B) + (DIST -| NULL) + (D11|-NULL)$)}{$A_{i+\frac{1}{2},j}$};
    \dimline[extension start length=-0.5cm, extension end length=-0.5cm,extension style={black}, label style={below=0.5ex}] {($(NULL-|B) + (DIST |- NULL)$)}{($(NULL-|B) + (DIST|- NULL) + (D11-|NULL)$)}{$A_{i+\frac{1}{2},j}$};
    
    \node[below left, xshift=-0.8cm, yshift=-0.8cm] at (C) {$\alpha_{i,j}$};

    \node[above right] at (NULL) {$\phi > 0$};
    \node[below left] at (B) {$\phi < 0$};


    \draw[->, line width=1pt] (-1,-1) -- (-1,0) node[left] {$y$};
    \draw[->, line width=1pt] (-1,-1) -- (0,-1) node[below] {$x$};

\end{tikzpicture}
    \caption{Schematic finite-volume discretization on a 2D Cartesian grid. The interface and its linear approximation are 
    represented by the red and blue line, respectively. The blue dash dotted line indicates the linear reconstruction on the subcell grid. The red dots represent the cell centers. The fluid within the cut cell corresponding to the positive and negative level-set
    values are represented by a grey and white background, respectively. Note that the volume fraction and apertures are computed for the positive fluid.
    (For interpretation of the references to color in this figure legend, the reader is referred to the web version of this article.)}
    \label{cut cell}
\end{figure}
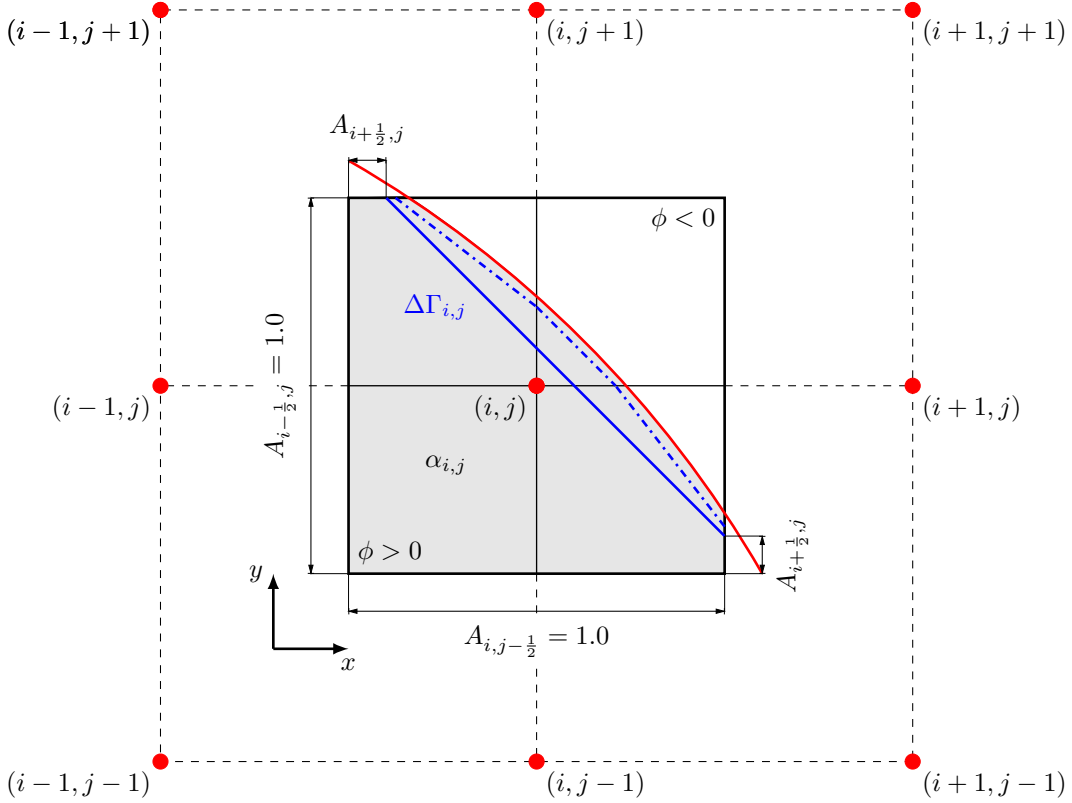

The level-set field $\phi$ represents the signed distance from the interface of each cell center within the mesh of the finite-volume discretization.
This implies that there is a positive phase ($\phi > 0$) and a negative phase ($\phi < 0$) with the interface being located at the zero level-set of $\phi$.
We normalize the level-set values with the cell size. A cell that is intersected by the interface is referred to as cut cell.
In Figure \ref{cut cell}, a schematic of the finite-volume discretization on a 2D Cartesian grid including a cut cell with circular interface is shown.

In the following, we describe the conventional linear interface reconstruction.
We then explain fundamentals of multilayer perceptrons and convolutional neural networks.
Finally, we introduce the novel data-driven interface reconstruction model.

\subsection{Linear interface reconstruction}
The level-set field is assumed to have a linear distribution along the cell faces of the cut cell. The first step in reconstructing the interface is to approximate the level-set 
values at the corners of the cut cell via bilinear interpolation from the level-set input $\bm{\phi}_{i,j} = \left[\phi_{i-1,j+1}, \phi_{i,j+1}, \dots, \phi_{i,j}, \dots, \phi_{i+1,j-1} \right] \in \mathbb{R}^9$. The corner values for the cell $(i,j)$ are given by
\begin{align}
    \begin{split}
        \phi_{i+\frac{1}{2},j+\frac{1}{2}} &= \frac{1}{4}(\phi_{i,j}+\phi_{i+1,j}+\phi_{i,j+1}+\phi_{i+1,j+1}) \\
        \phi_{i+\frac{1}{2},j-\frac{1}{2}} &= \frac{1}{4}(\phi_{i,j}+\phi_{i+1,j}+\phi_{i,j-1}+\phi_{i+1,j-1}) \\
        \phi_{i-\frac{1}{2},j+\frac{1}{2}} &= \frac{1}{4}(\phi_{i,j}+\phi_{i-1,j}+\phi_{i,j+1}+\phi_{i-1,j+1}) \\
        \phi_{i-\frac{1}{2},j-\frac{1}{2}} &= \frac{1}{4}(\phi_{i,j}+\phi_{i-1,j}+\phi_{i,j-1}+\phi_{i-1,j-1}). 
    \end{split}
\end{align}
If the level-set function is changing its sign along a cell face, the cell face is cut by the interface. In this case, the apertures $\bm{A}_{i,j} = \left[ A_{i,j+\frac{1}{2}}, A_{i,j-\frac{1}{2}}, A_{i+\frac{1}{2},j}, A_{i-\frac{1}{2},j} \right] \in \mathbb{R}^4$  
are computed as
\begin{align}
    \begin{split}
        A_{i,j+\frac{1}{2}}=b_{i,j+\frac{1}{2}} \ \ \ \text{if} \ \ \ \phi_{i+\frac{1}{2},j+\frac{1}{2}} > 0 \ \ \ \text{else} \ \ \ 1 - b_{i,j+\frac{1}{2}}, \\
        A_{i,j-\frac{1}{2}}=b_{i,j-\frac{1}{2}} \ \ \ \text{if} \ \ \ \phi_{i+\frac{1}{2},j-\frac{1}{2}} > 0 \ \ \ \text{else} \ \ \ 1 - b_{i,j-\frac{1}{2}}, \\
        A_{i+\frac{1}{2},j}=a_{i+\frac{1}{2},j} \ \ \ \text{if} \ \ \ \phi_{i+\frac{1}{2},j+\frac{1}{2}} > 0 \ \ \ \text{else} \ \ \ 1 - a_{i+\frac{1}{2},j}, \\
        A_{i-\frac{1}{2},j}=a_{i-\frac{1}{2},j} \ \ \ \text{if} \ \ \ \phi_{i-\frac{1}{2},j+\frac{1}{2}} > 0 \ \ \ \text{else} \ \ \ 1 - a_{i-\frac{1}{2},j}, \\
    \end{split}
\end{align}
with
\begin{align}
    \begin{split}
        a_{i\pm\frac{1}{2},j} = \frac{|\phi_{i\pm\frac{1}{2},j+\frac{1}{2}}|}{|\phi_{i\pm\frac{1}{2},j+\frac{1}{2}}| + |\phi_{i\pm\frac{1}{2},j-\frac{1}{2}}|}, \\
        b_{i,j\pm\frac{1}{2}} = \frac{|\phi_{i+\frac{1}{2},j\pm\frac{1}{2}}|}{|\phi_{i+\frac{1}{2},j\pm\frac{1}{2}}| + |\phi_{i-\frac{1}{2},j\pm\frac{1}{2}}|}. \\
    \end{split}
\end{align}
If there is no sign change along the cell face, the aperture is either 0 or 1, depending on the level-set sign at the cut cell center. Note that both the apertures and 
the volume fraction are computed according to the positive fluid ($\phi > 0$).

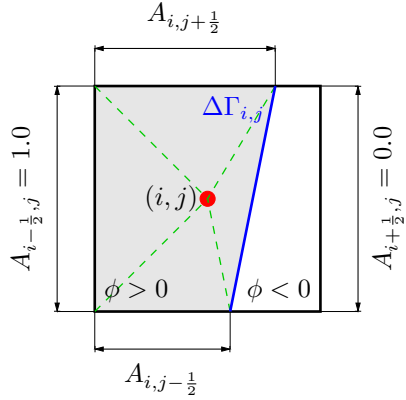
\begin{figure}[!t]
    \centering
    \begin{tikzpicture}
    \coordinate (NULL) at (0,0);
    \coordinate (A) at (3,3);
    \coordinate (B1) at ($0.8*(NULL -| A)$);
    \coordinate (B2) at ($0.6*(NULL -| A)$);

    \draw[fill=black!10] (NULL) -- (0,3) -- (A-|B1) -- (NULL-|B2) -- cycle;

    \draw[line width=1pt] (NULL) rectangle (A);
    \draw[fill=red, draw=red] ($0.5*(A)$) circle (0.1) node[left] {$(i, j)$};

    \draw[line width=0.5pt, green!80!black, dashed] ($(NULL |- A) + (B1)$) -- ($0.5*(A)$);
    \draw[line width=0.5pt, green!80!black, dashed] ($(NULL |- A)$) -- ($0.5*(A)$);
    \draw[line width=0.5pt, green!80!black, dashed] (NULL) -- ($0.5*(A)$);
    \draw[line width=0.5pt, green!80!black, dashed] (B2) -- ($0.5*(A)$);

    \draw[line width=1pt, blue] ($(NULL |- A) + (B1)$) -- (B2) node[at start, below left] {$\Delta \Gamma_{i,j}$};

    \node[above right] at (0,0) {$\phi>0$};
    \node[above left] at (3,0) {$\phi<0$};

    \coordinate (DIST) at (0.5,0.5);
    \dimline[extension start length=0.5cm, extension end length=0.5cm,extension style={black}, label style={above=0.5ex}] {($(NULL|-A) + (NULL|-DIST)$)}{($(NULL |- A) + (B1) + (NULL|-DIST)$)}{$A_{i,j+\frac{1}{2}}$};
    \dimline[extension start length=-0.5cm, extension end length=-0.5cm,extension style={black}, label style={below=0.5ex}] {($-1*(NULL|-DIST)$)}{($(B2) - (NULL|-DIST)$)}{$A_{i,j-\frac{1}{2}}$};
    \dimline[extension start length=0.5cm, extension end length=0.5cm,extension style={black}, label style={above=0.5ex}] {($-1*(NULL-|DIST)$)}{($(NULL|-A) - (NULL-|DIST)$)}{$A_{i-\frac{1}{2},j}=1.0$};
    \dimline[extension start length=-0.5cm, extension end length=-0.5cm,extension style={black}, label style={below=0.5ex}] {($(NULL-|A) + (NULL-|DIST)$)}{($(A) + (NULL-|DIST)$)}{$A_{i+\frac{1}{2},j}=0.0$};
 
\end{tikzpicture}
    \caption{Schematic of the volume fraction computation in the linear interface reconstruction.}
    \label{LIR_volume_fraction}
\end{figure}

The interface segment-length $\Delta \Gamma_{i,j}$ and the volume fraction $\alpha_{i,j}$ are evaluated as
\begin{align}
    \Delta\Gamma_{i,j} &= \left[(A_{i+\frac{1}{2},j} - A_{i-\frac{1}{2},j})^2 + (A_{i,j+\frac{1}{2}} - A_{i,j-\frac{1}{2}})^2 \right]^\frac{1}{2} \\
    \alpha_{i,j} &= \frac{1}{2}\left( \frac{1}{2}A_{i+\frac{1}{2},j} + \frac{1}{2}A_{i-\frac{1}{2},j} + \frac{1}{2}A_{i,j+\frac{1}{2}} + \frac{1}{2}A_{i,j-\frac{1}{2}} + \Delta\Gamma_{i,j}\phi_{i,j} \right)
    \label{volume_fraction_computation_linear}
\end{align}
As depicted in Figure \ref{LIR_volume_fraction}, the volume fraction is the sum of the area of triangles that meet at the cell center.

To increase the accuracy of the described linear interface reconstruction, Luo et al. \cite{Luo2015b} proposed a subcell-resolution interface reconstruction scheme.
Here, the cut cell is refined once into four cells. Subsequently, the above described scheme is applied to each subcell. The final volume fraction and apertures 
are then computed by averaging the corresponding subcell apertures and volume fractions. The subcell reconstruction is used in this work.

\subsection{Deep learning models}

\subsubsection{Multilayer perceptron}
Neural networks are parameterizable nonlinear compound functions that map an input vector $\bm{x}$ onto an output $\bm{y} = \mathcal{N}(\bm{x}, \bm{\theta})$,
where the free and learnable parameters of the network $\bm{\theta} = \left\lbrace \bm{W}^l, \bm{b}^l \right\rbrace$ are comprised of the weights and biases of the network layers.
Deep neural networks (DNNs) consist of multiple hidden layers of units (neurons) between in- and output layer.
The numerical values in each layer are called activations.

The basic neural network model is the multilayer peceptron (MLP), where neurons in adjacent layers are densely connected.
The vector of activations $\bm{a}^{l}$ in layer $l$ is computed from the activations of the previous layer $\bm{a}^{l-1}$ as

\begin{align}
    \bm{a}^{l} &= \sigma\left( \bm{W}^{l-1} \bm{a}^{l-1} + \bm{b}^{l-1} \right), 
    \label{eq:DNN}
\end{align}
where $\bm{W}^{l-1}$ indicates the weight matrix linking layers $(l-1)$ and $l$, $\bm{b}^{l-1}$ is the bias vector, and $\sigma()$ is an elementwise nonlinearity.
A suitable set of parameters $\bm{\theta}$ is found by approximately minimizing the error $\mathcal{L}(\bm{y}, \hat{\bm{y}})$ between network output $\bm{y}$ and target output $\hat{\bm{y}}$, e.g. mean squared error.
Typically, the loss function is minimized via mini-batch gradient descent or the popular Adam optimizer \cite{Kingma2015}.\\

\subsubsection{Convolutional neural network}
Convolutional neural networks (CNNs) are a special type of deep neural network for structured input data.
CNNs exploit sparsity and weight sharing to increase learning efficiency.
First, neuron activations are computed only by local sparse interactions with neurons from the previous layer.
The local weight matrix is called the filter kernel.
Second, CNNs make use of parameter sharing, i.e. the same weight matrix is used for each neuron in a given layer, and, therefore, are translational invariant.
Each CNN layer is typically made out of multiple feature maps, each of which is computed by a corresponding filter.



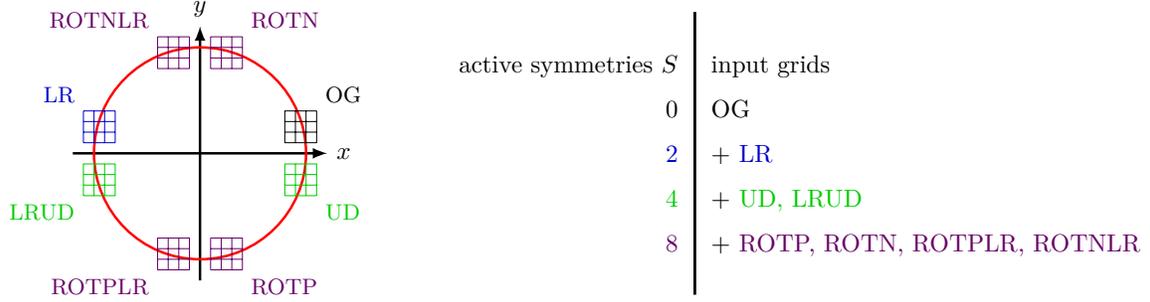
\begin{figure}[t]
    \centering
    \adjustbox{max width=\textwidth}{
    \begin{tikzpicture}
    \coordinate (NULL) at (0,0);
    \coordinate (W) at (1.8,1.8);

    \draw[->, line width=1pt] ($-1*(NULL |- W)$) -- ($(NULL |- W)$) node[at end, above] {$y$};
    \draw[->, line width=1pt] ($-1*(NULL -| W)$) -- ($(NULL -| W)$) node[at end, right] {$x$};
    \draw[line width=1.0pt, red] (NULL) circle (1.5cm);

    \draw[step=0.15cm] (1.199,0.15) grid (1.65,0.6) node[above right] {\small OG};
    \draw[step=0.15cm, blue!80!black] (-1.199,0.15) grid (-1.65,0.6) node[above left] {\small LR};
    \draw[step=0.15cm, green!80!black] (1.199,-0.15) grid (1.65,-0.6) node[below right] {\small UD};
    \draw[step=0.15cm, green!80!black] (-1.199,-0.15) grid (-1.65,-0.6) node[below left] {\small LRUD};

    \draw[step=0.15cm, violet!80!black] (0.15,1.199) grid (0.6,1.65) node[above right] {\small ROTN};
    \draw[step=0.15cm, violet!80!black] (-0.15,1.199) grid (-0.6,1.65) node[above left] {\small ROTNLR};
    \draw[step=0.15cm, violet!80!black] (0.15,-1.199) grid (0.6,-1.65) node[below right] {\small ROTP};
    \draw[step=0.15cm, violet!80!black] (-0.15,-1.199) grid (-0.6,-1.65) node[below left] {\small ROTPLR};

    \draw[line width=1pt] (7,2) -- (7,-2) node[left, left, midway, xshift=-0.1cm, align=right] {active symmetries $S$ \\ 0 \\ \color{blue!80!black} 2 \\ \color{green!80!black} 4 \\ \color{violet!80!black} 8};
    \draw[line width=1pt] (7,2) -- (7,-2) node[left, right, midway, yshift=-0.035cm, xshift=0.1cm, align=left] {input grids \\ OG \\ + \color{blue!80!black} LR \\ + \color{green!80!black} UD, LRUD \\ + \color{violet!80!black} ROTP, ROTN, ROTPLR, ROTNLR};

\end{tikzpicture}
    }
    \caption{(Left) Schematic of all used symmetries. The input level-set grid (OG) is mirrored across the $y$-axis (LR), across the $x$-axis (UD) and both (LRUD).
    Furthermore, the input level-set grid is rotated $90^\circ$ clockwise (ROTP), anti-clockwise (ROTN) and both resulting grids are mirrored across the $y$-axis
    (ROTPLR and ROTNLR). (Right) Naming convention for active symmetries in downstream task. For example, $S=2$ means left-right symmetry (LR) is being used.}
    \label{symmetries_fig}
\end{figure}

\begin{figure}[!b]
    \centering
    \input{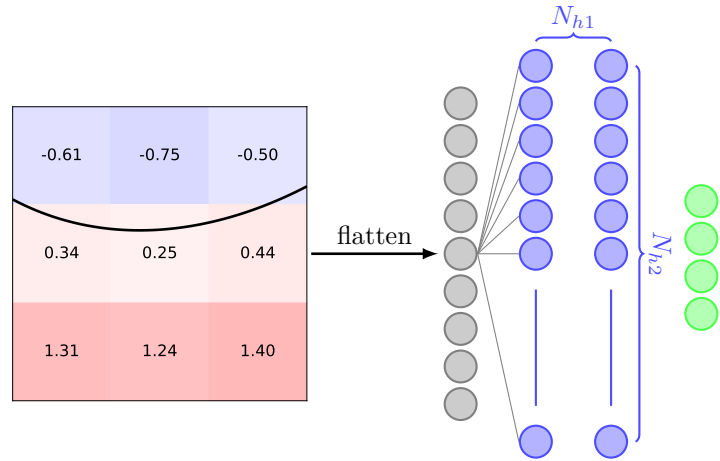}
    \caption{Schematic of a multilayer perceptron to predict apertures from $3\times3$ level-set input. The aperture neural network has $4$ output nodes. For the volume fraction, the output layer depicted in green consists of a single node.}    
    \label{mlp}
\end{figure}

\begin{figure}[t]
    \centering
    \input{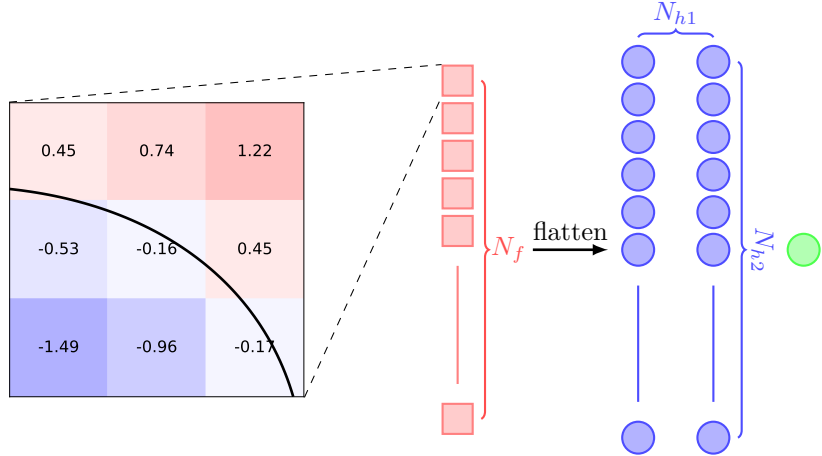}
    \caption{Schematic of a convolutional neural network to predict volume fraction from $3\times3$ level-set input. For the apertures, the output layer depicted in green consists of 4 nodes. For the classification task, the output layer consists of two nodes.}    
    \label{cnn}
\end{figure}

\subsection{Combined model}

\label{section:combined_model}
The data-driven combined interface model uses neural networks for the calculation of the volume fraction and the apertures of cell $(i,j )$, i.e. 
\begin{align}
    \alpha_{i,j} &= \mathcal{N}_{\alpha}(\bm{\phi}_{i,j}, \bm{\theta}_{\alpha}), \\
    \bm{A}_{i,j} &= \mathcal{N}_{A}(\bm{\phi}_{i,j}, \bm{\theta}_{A}), 
\end{align}
where $\mathcal{N}_{\alpha}$ is the volume fraction neural network and $\mathcal{N}_{A}$ the aperture neural network, respectively.

We pose two requirements with respect to the data-driven interface reconstruction model:
Firstly, we want the model to be equivariant under symmetric transformations (e.g. reflection of the input).
Secondly, we want the model to converge upon mesh refinement.

We preserve symmetry at the additional cost of multiple neural network evaluations by mirroring and rotating the input level-set grid and subsequently averaging the predictions.
The left half of Fig. \ref{symmetries_fig} shows a schematic of the implemented symmetries on the example of a circular interface.
Starting from the the original configuration (OG), up to $7$ symmetric and rotational variations are generated.

In practice, the absolute error of neural networks can be significantly smaller than that of classical discrete schemes for certain resolution ranges.
However, the asymptotic error behavior of neural networks models is often limited by the level of the generalization error, and convergence upon mesh refinment is not guaranteed.
Therefore, we propose to use a hybrid approach in which we use a neural network interface reconstruction for coarse resolutions while switching to the classical linear reconstruction in the asymptotic limit.
A second smaller neural network works as a classifier and decides based on the input level-set field whether the interface reconstruction should be handled by the conventional method or by the neural network reconstruction.

\begin{figure}[t]
    \centering
    \adjustbox{max width=\textwidth}{
    \begin{tikzpicture}
    \coordinate (NULL) at (0,0);
    \coordinate (RIGHT) at (2.5,0);
    \coordinate (DOWN) at (0,-2.5);
    \coordinate (A) at (0,0);
    \coordinate (B1) at (2.6,0);
    \coordinate (B) at ($(A) + 2*(B1)$);
    \coordinate (C1) at ($(B) + (3.1,1.7)$);
    \coordinate (C2) at ($(B) + (3.1,-1.7)$);
    \coordinate (P1) at ($0.48*(B1)$);
    \coordinate (P2) at ($(B) + (3.3,0)$);
    \coordinate (P3) at ($0.5*(B) + 0.5*(B1) - (0.15,0)$);
    \coordinate (D1) at ($(C1) + (3.0,0.0)$);
    \coordinate (D2) at ($(C2) + (2.5,0.0)$);
    \coordinate (E) at ($(D1) + (2.0,-1.25)$);

    \node[draw=black, line width=1pt, rounded corners=0.1cm, align=center, execute at begin node=\setlength{\baselineskip}{1em}] (A1) at (0,0) {Input levelset \\ grid $\bm{\phi}_{i,j}$};
    \node[draw=black, line width=1pt, rounded corners=0.1cm, align=center, execute at begin node=\setlength{\baselineskip}{1em}] (B11) [right = 0.3cm of A1] {Generate \\ symmetries};
    \node[draw=black, line width=1pt, rounded corners=0.1cm] (B1) [right = 0.3cm of B11] {Classifier $\epsilon$};
    \node[draw=black, line width=1pt, rounded corners=0.1cm, align=center, execute at begin node=\setlength{\baselineskip}{1em}] (B111) [right = 0.3cm of B1] {Average \\ symmetries};
    \node[diamond, draw=black, line width=1pt, inner sep = 0.1cm] (IF) [right = 0.3cm of B111] {if};
    \node[draw=black, line width=1pt, rounded corners=0.1cm, align=center, execute at begin node=\setlength{\baselineskip}{1em}] (C11) [above = 1cm of IF] {Neural network \\ reconstruction};
    \node[draw=black, line width=1pt, rounded corners=0.1cm, align=center, execute at begin node=\setlength{\baselineskip}{1em}] (C22) [below = 1cm of IF] {Linear \\ reconstruction};
    \node[draw=black, line width=1pt, rounded corners=0.1cm, align=center, execute at begin node=\setlength{\baselineskip}{1em}] (D11) [right = 0.3cm of C11] {Average \\ symmetries};
    \node[draw=black, line width=1pt, rounded corners=0.1cm] (E1) [below right = 1cm and 0.3cm of D11] {$\alpha_{i,j}, \bm{A}_{i,j}$};

    \draw[->, line width=1pt] (A1.east) -- (B11.west);
    \draw[->, line width=1pt] (B11.east) -- (B1.west);
    \draw[->, line width=1pt] (B1.east) -- (B111.west);
    \draw[line width=1pt] (B111.east) -- (IF);
    
    \draw[<-, line width=1pt] (C11.south) -- (IF.north) node[right, midway] {$\epsilon_{nn} > 0.3$};
    \draw[<-, line width=1pt] (C22.north) -- (IF.south) node[right, midway] {$\epsilon_{nn} < 0.3$};;

    \draw[->, line width=1pt] (C11.east) -- (D11.west);
    \draw[->, line width=1pt] (D11.east) to [out=0, in=90, distance=0.8cm] (E1.north);
    \draw[->, line width=1pt] (C22.east) to [out=0, in=-90, distance=1.8cm] (E1.south);

\end{tikzpicture}
    }
    \caption{Schematic of the combined model.}
    \label{combined}
\end{figure}
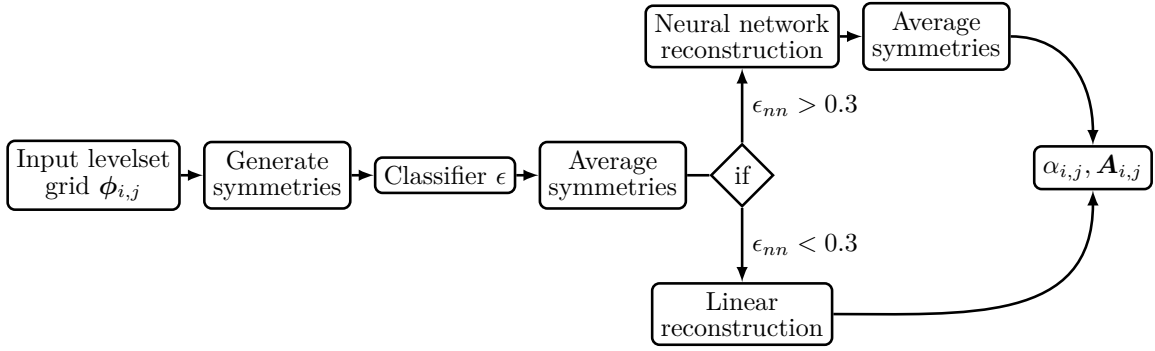

The whole algorithm is illustrated in Fig. \ref{combined}.
First, all user-specified symmetric/rotational variations of the level set input are generated.  
Based on the properties of the input field, the classifier $\epsilon$ decides whether the linear or the data-driven reconstruction are to be preferred.
In particular, we choose a classifier probability threshold $\epsilon_{nn} > 0.3$ for all presented tests in the following sections.
The linear reconstruction is inherently equivariant to symmetries and can be evaluated directly.
The neural network is evaluated for each variation of the original input, the results are then averaged. 

All neural networks - the classifier. the volume fraction and aperture model - are trained outside of the CFD solver on generic interface configurations.
Details of the training process are provided in \ref{sec:ModelTraining}.
In the ALPACA CFD solver the reconstruction of the volume fraction and the apertures is computed in different parts of the code.
To minimize computational overhead, we train a separate reconstruction network for each geometric quantity.
A single model for multiple quantities would require a deeper neural network, as the prediction task is more complex, and would result in a worse performance as only one part of the output would be used.

In previous work, only multilayer perceptrons were used to perform the regression task. 
Here, we are also training convolutional neural networks to investigate whether the regression task benefits from 
their capability to account for the spatial structure of the level-set grid.
The linear interface reconstruction utilizes a $3\times 3$ level-set grid as input to compute volume fraction and apertures.
To make both reconstructions comparable, we use the same stencil width as input for the regression neural networks.
For the classification task, we only consider convolutional type networks.
However, for the classification neural network, we are using input grids of size $3\times 3$ and $5\times 5$.  

In Fig. \ref{mlp}, the architecture of the MLP is illustrated. 
The $3\times 3$ level-set grid is flattened and used as input for the network. 
The amount of hidden layers and the amount of nodes per hidden layer is represented by $N_{h1}$ and $N_{h2}$, respectively. 
Depending on the model, the output layer consists of one (volume fraction) or four (apertures) nodes. The naming convention to 
describe the MLP architecture in this work is \textit{NumberOfLayers NumberOfNodesPerLayer}. As an example, the model \textit{2 50} 
describes an MLP consisting of 2 hidden layers with 50 nodes each.

The architecture of the convolutional neural network is displayed in Fig. \ref{cnn}. 
The $3\times 3$ (or $5\times 5$) level-set grid is processed by a single convolutional layer with $N_f$ filters of size $3$ (or $5$).
The output of the convolutional layer is flattened to a vector and used as input for $N_{h1}$ subsequent dense layers with $N_{h2}$ nodes each. 
Depending on the model, the output layer consists of one (volume fraction) or four (apertures) nodes. 
For the classifier, the output layer contains a node for each class, i.e. two nodes.
The naming convention to describe the CNN architecture in this work is \textit{KernelWidth NumberOfFilters NumberOfDenseLayers NumberOfNodesPerLayer}. As an example, the model \textit{3 16 2 20} 
describes a CNN consisting of a convolutional layer composed of 16 filters, each with a kernel width of 3 and 2 subsequent dense layers with 20 nodes each.

\section{Data generation and datasets} 
\label{data generation}

The ML models are trained using generic interfaces described by lines, circles and ellipses. 
Furthermore, the test set contains star shaped interfaces, i.e. circles whose radius has a sinusoidal dependency.
The parametric representation $\mathbf{\Xi}(\theta)$ of the circles, ellipses and stars are given by
\begin{align}
    \begin{pmatrix}
        x \\
        y
    \end{pmatrix}
    = \ \mathbf{\Xi}_c(\theta)&=
    \begin{pmatrix}
        R_c\cos(\theta) \\
        R_c\sin(\theta) 
    \end{pmatrix} &&\text{circles} \\
    \mathbf{\Xi}_e(\theta)&=
    \begin{pmatrix}
        a_e\cos(\theta)\cos(\psi) - b_e\sin(\theta)\sin(\psi) \\
        a_e\cos(\theta)\sin(\psi) + b_e\sin(\theta)\cos(\psi) 
    \end{pmatrix}  &&\text{ellipses} \nonumber \\
    \mathbf{\Xi}_s(\theta)&=
    \begin{pmatrix}
        (R_s + A_s\cos(n_s\theta))\cos(\theta)\cos(\psi) - (R_s + A_s\cos(n_s\theta))\sin(\theta)\sin(\psi) \\
        (R_s + A_s\cos(n_s\theta))\cos(\theta)\sin(\psi) + (R_s + A_s\cos(n_s\theta))\sin(\theta)\cos(\psi) 
    \end{pmatrix} &&\text{stars} \nonumber
\end{align}
Here, $R_c$ represents the radius of the circles and $a_e$ and $b_e$ are the semi-major and semi-minor of the ellipses, respectively. The stars are described by the base radius $R_s$,
the amplitude $A_s$ and the number of fingers $n_s$, respectively. The parameter $\theta$ describes the angle along the curve perimeter and $\psi$ represents a rotation angle.
Note that all parameters that represent length scales are given as normalized with the cell size, i.e. larger radii indicate higher numerical resolution.

\setlength\arrayrulewidth{1pt}
\def\arraystretch{1.0}
\begin{table}[b]
    \centering
    \adjustbox{max width=\textwidth}{
    \begin{tabular}{  p{1.2cm} | p{0.7cm} p{3.1cm} p{4.0cm} p{6.0cm} }
        dataset & lines & circles & ellipses & stars \\
        \hline
        \hline
        \textit{1} & 400 & 400 $R_c\in[2,200]$ & 200 $(a_e,b_e)\in[2,5]\times[2,20]$  & - \\
        \textit{2} & 200 & 200 $R_c\in[2,200]$ \newline 200 $R_c\in[200,4000]$ & 100 $(a_e,b_e)\in[2,5]\times[2,20]$ & - \\
        \textit{3} & 10 & 10 $R_c\in[2,200]$ & 10 $(a_e,b_e)\in[2,5]\times[2,20]$ & 10 $(R_s,A_s,n_s)\in[4,20]\times[1.2,14]\times[3,7]$\\
    \end{tabular}
    }
    \caption{Summary of datasets. The amount of samples are given in 1000. For interpretation}
    \label{datasets_table}
\end{table}

The general approach to generate a sample for any given parametric representation is as follows:
\begin{enumerate}
    \item Define a 2D Cartesian coordinate system with origin $(0,0)$ used as center of the parametric curve.
    \item Draw parameters describing the shape $(R_c,a_e,b_e,R_s,A_s,n_s)$ of the curve from a random uniform distribution. Pick any cut cell along the perimeter by drawing $\theta_0$ from a random uniform distribution.
    \item Identify the cut cell center by rounding $\mathbf{\Xi}(\theta_0)$ and subsequently generate a $3\times 3$ (or $5\times5$) grid of cells surrounding the cut cell, with $\mathbf{x}^\phi_{i,j}, \forall i,j = \{1,2,3\}$ being the cell centers of the grid.
    \item Compute the levelset field $\bm{\phi}_{i,j}$: The distance of the cell centers $\mathbf{x}^\phi$ to the interface is computed by finding the minimum of $||\mathbf{x}^\phi - \mathbf{\Xi}(\theta)||$ using the Newton method. To infer the sign of the levelset value, the implicit form of the curve is used for circles ($0=R_c - \sqrt{x^2+y^2}$) and ellipses ($0=1 - \sqrt{ \sfrac{(x\cos(\psi)+y\sin(\psi))^2}{a_e^2} + \sfrac{(x\sin(\psi)-y\cos(\psi))^2}{b_e^2} }$). For stars, the radius at the cell centers $||\mathbf{x}_\phi||$ is compared to the radius of the curve at the corresponding angle. 
    \item Compute the apertures $\bm{A}_{i,j}$: Find the intersections of the parametric curve with all cell faces of the cut cell using the Newton method and distinguish between positive and negative fluid using the normal.
    \item Compute the volume fraction $\alpha$: Analytical integration for circles and numerical integration (Simpsons rule) for ellipses and stars.
    \item Decide whether to flip the sample from convex to concave, i.e. multiply the levelset field with -1, by drawing from a random uniform distribution.
\end{enumerate}

\begin{figure}[t]
    \centering
    \begin{subfigure}[b]{0.24\textwidth}
        \resizebox{35mm}{!}{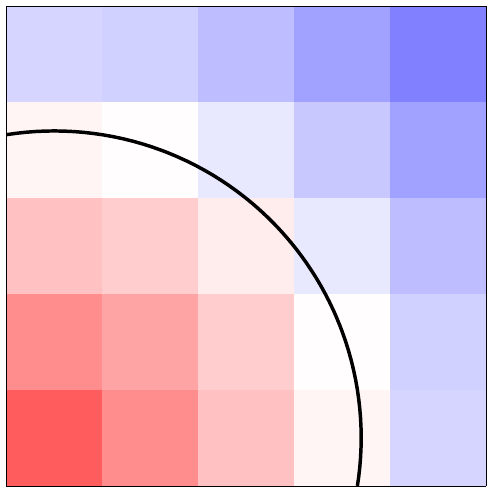}
    \end{subfigure}
    \begin{subfigure}[b]{0.24\textwidth}
        \resizebox{35mm}{!}{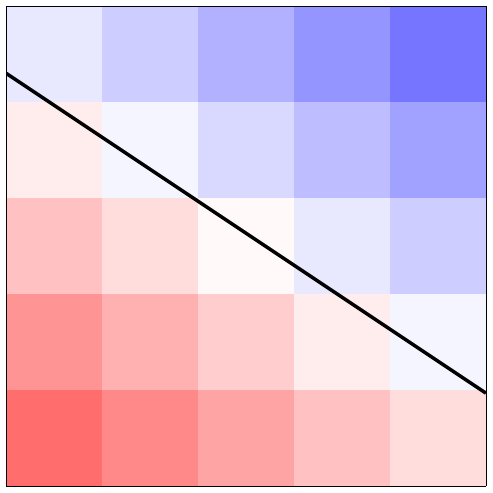}
    \end{subfigure}
    \begin{subfigure}[b]{0.24\textwidth}
        \resizebox{35mm}{!}{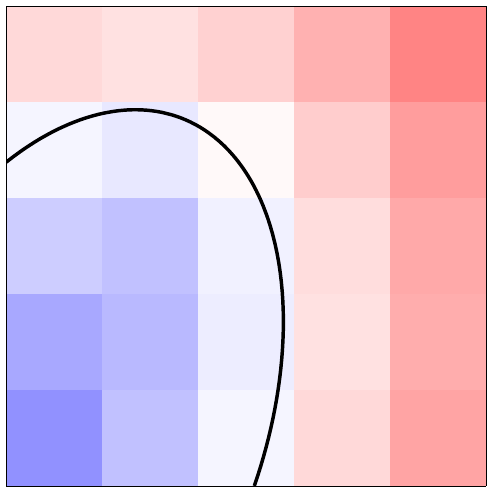}
    \end{subfigure}
    \begin{subfigure}[b]{0.24\textwidth}
        \resizebox{35mm}{!}{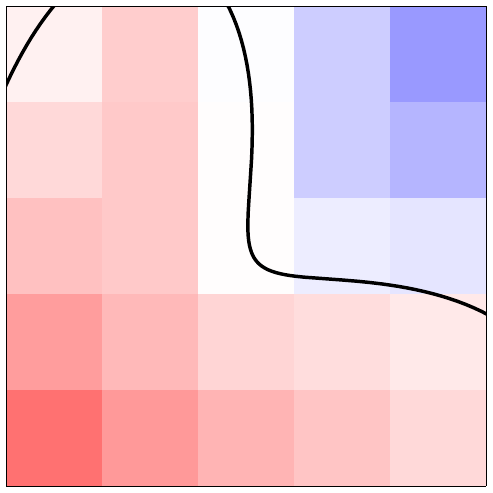}
    \end{subfigure}

    \caption{Exemplary levelset grids for all used shapes. From left to right: Circle, line, ellipse, and star.}
    \label{shapes}
\end{figure}

For lines, the cut cell center is placed at the origin of the 2D Cartesian coordinate system. Subsequently, two points within the cut cell, i.e. $(x,y)\in[-0.5,0.5]$, are randomly picked from a uniform distribution.
The line is defined by connecting both points. The apertures are computed by finding the intersection of the line with the cut cell faces. The volume fraction is then computed according to \eqref{volume_fraction_computation_linear} (compare Figure \ref{LIR_volume_fraction}).

\begin{figure}[b]
    \centering
    \input{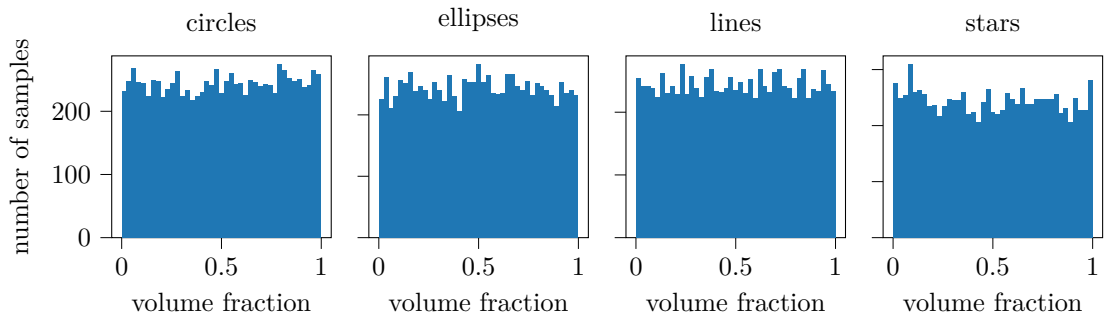}
    \caption{Histogram of volume fraction over samples in the test set (dataset 3 in Table \ref{datasets_table}).}
    \label{volume_fraction_histogram}
\end{figure}

Figure \ref{shapes} illustrates exemplary levelset grids for all considered shapes. In Figure \ref{volume_fraction_histogram}, a histogram of the volume fraction over all samples in the used test set 
is displayed. This shows, that the above described method yields a uniform distribution over labels preventing training data induced biases.

As best practiced in ML, we distinguish between training, validation and test set. The used datasets are summarized in Table \ref{datasets_table}. 
\textit{Dataset 1} and \textit{2} represent the training set for the regression and classification neural network, respectively, where $17.5\%$ of the samples are used as validation data.
\textit{Dataset 3} is the test set, that is evaluated after training for final model comparison and selection. Note, that the training sets are mirrored for all symmetries (compare Figure \ref{symmetries_fig}), i.e. the actual amount 
of samples is 8 times larger than what is described in the table. In addition, a fourth dataset is generated that is used to assess the convergence of the models. For this 
dataset, 4000 circles and 4000 ellipses are generated at each of the radii and semi-majors $R_c=a_e=10^r, r\in\{0.6,1.2,1.5,1.8,2.1,2.4,2.7,3.0,3.3,3.6,3.9\}$.
The semi-minor $b_e$ for each individual sample is computed by drawing an aspect ratio $a_e/b_e\in[2,4]$ from a random uniform distribution.

The labels of the training set for the classification neural networks are generated as follows:
For a given sample from the training set of the classification neural network (\textit{dataset 2}), we perform a neural network and linear interface reconstruction and subsequently compare the resulting
volume fractions and apertures to the analytical result. Depending on which model yields a lower absolute error, we generate a corresponding label for the given classifier training set sample.
This implies that every classification neural network model corresponds to a specific regression neural network model.

\section{Model performance} 
\label{model perfomance}

\subsection{Influence of active symmetries}

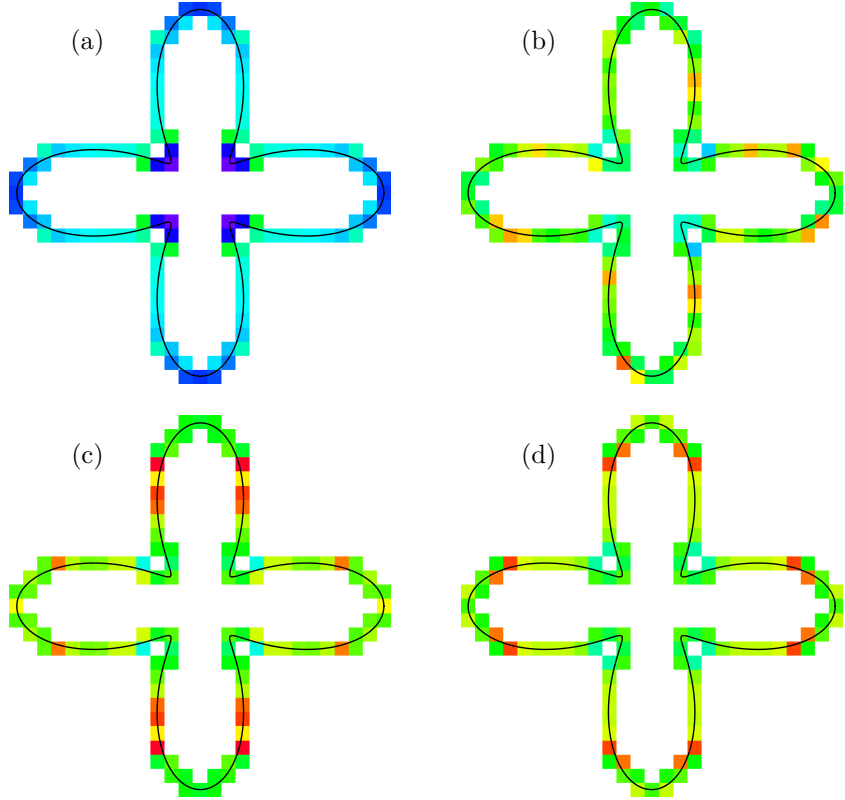
\begin{figure}[!b]
    \centering
    \begin{tikzpicture}
        \node at (0.0,0) {\includegraphics[scale=0.35]{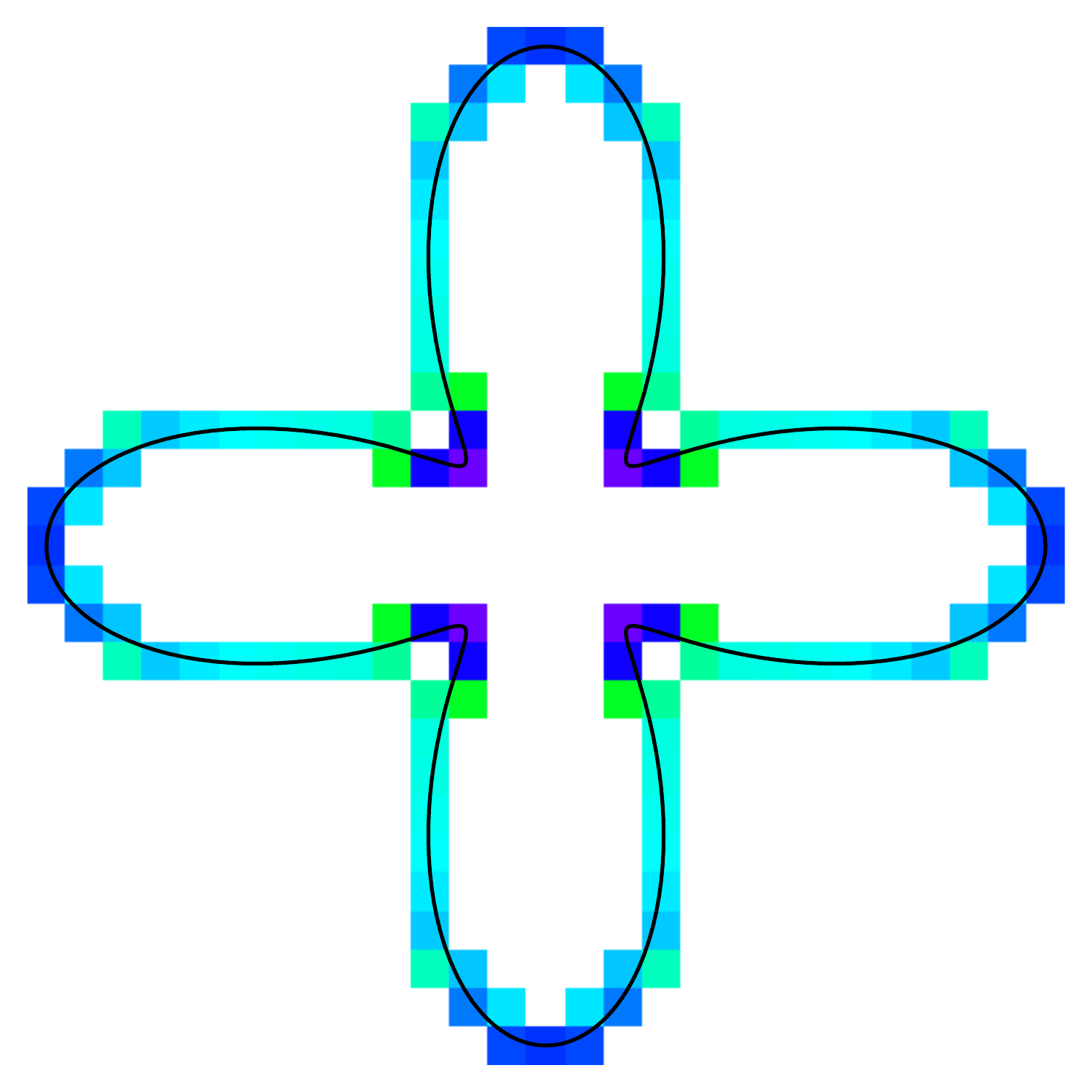}};
        \node at (-1.5,2) {(a)};
        \node at (6.0,0) {\includegraphics[scale=0.35]{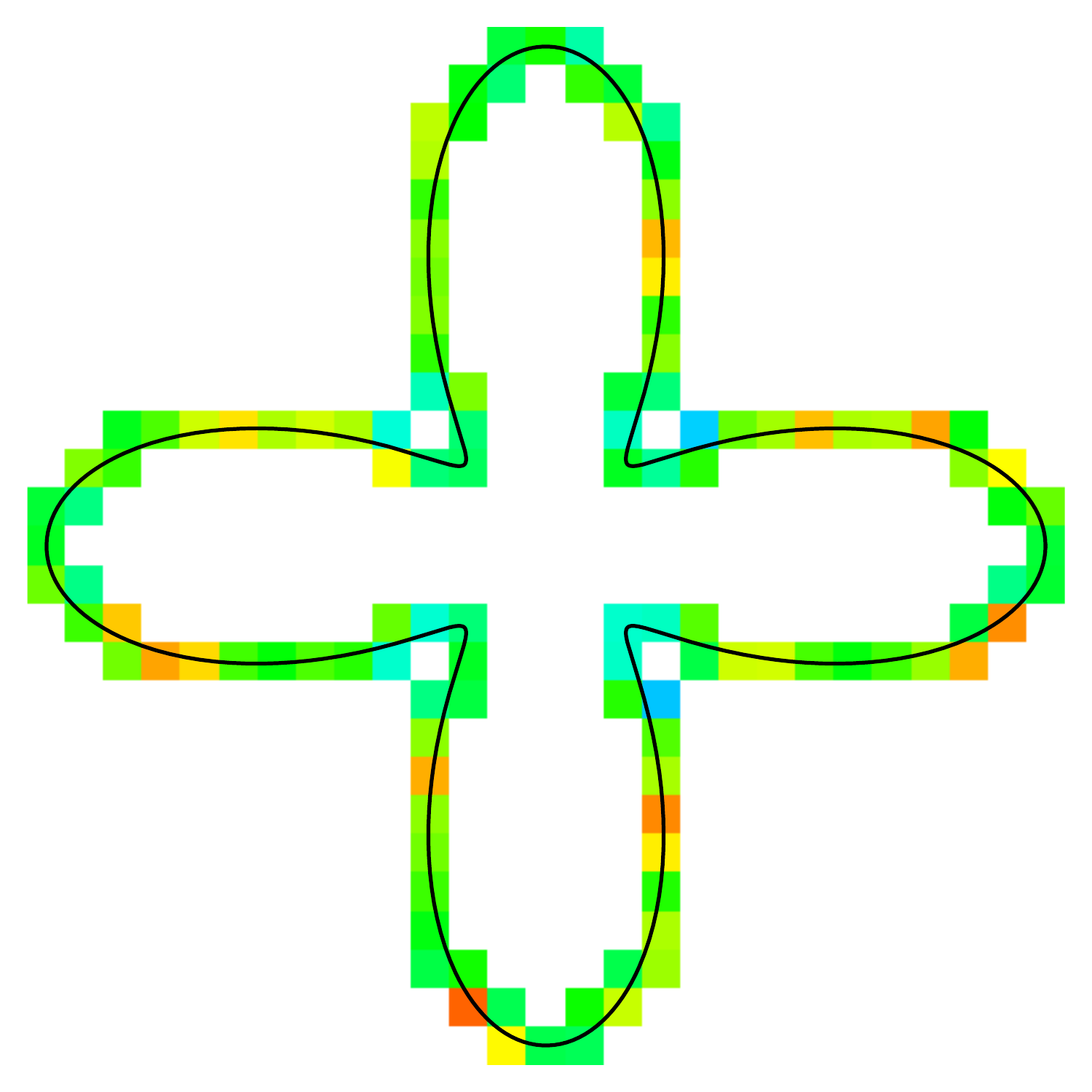}};
        \node at (4.5,2) {(b)};
        \node at (0.0,-5.5) {\includegraphics[scale=0.35]{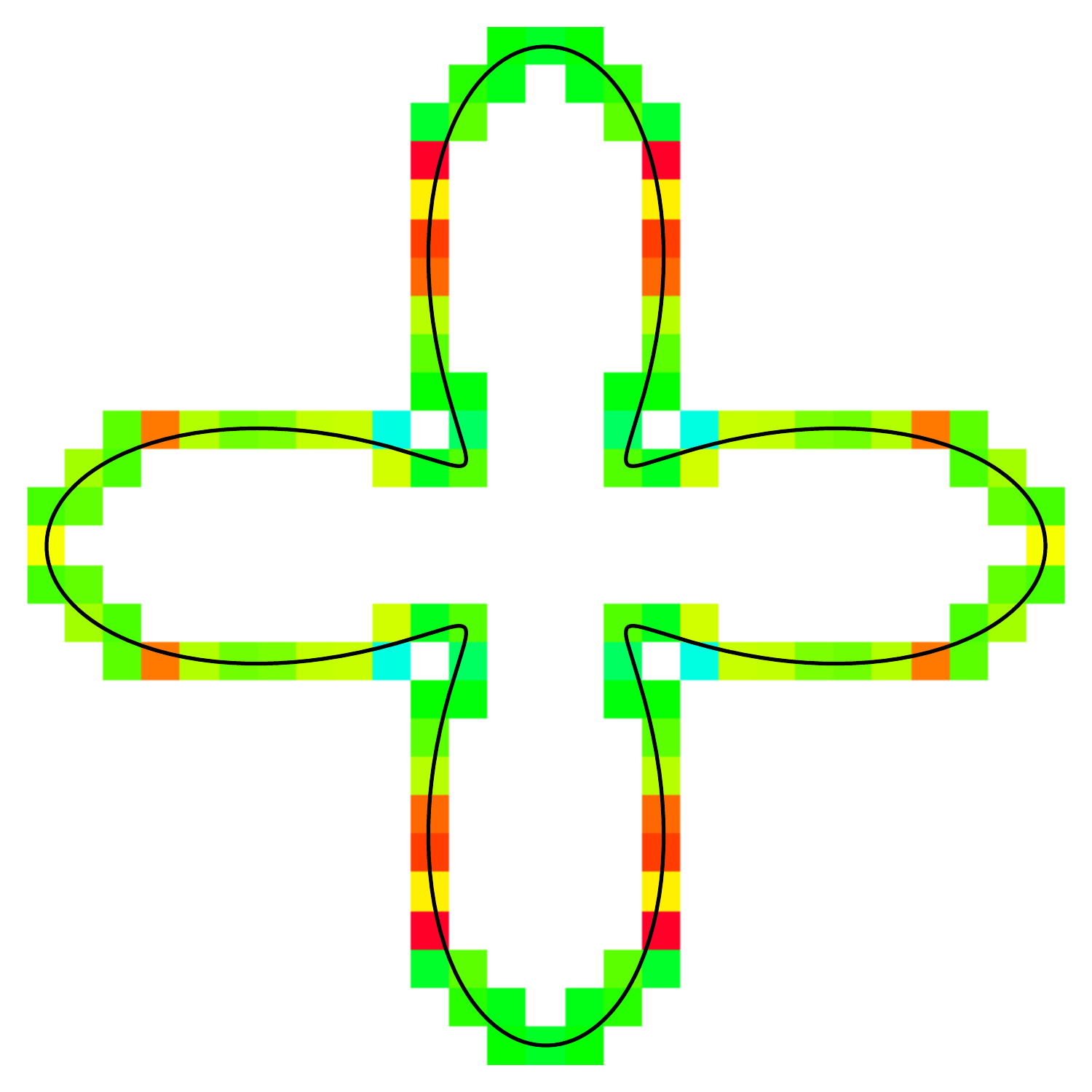}};
        \node at (-1.5,-3.5) {(c)};
        \node at (6.0,-5.5) {\includegraphics[scale=0.35]{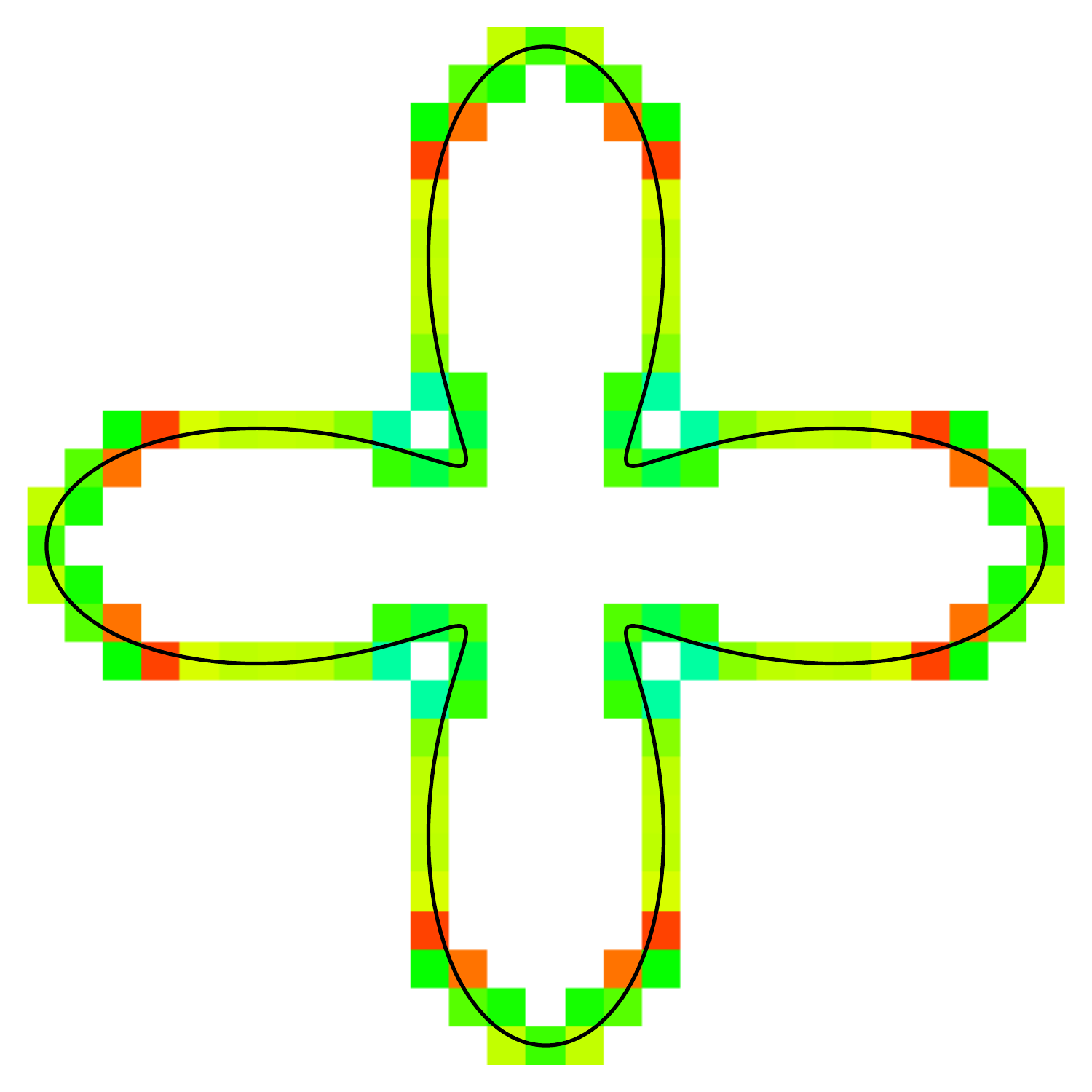}};
        \node at (4.5,-3.5) {(d)};
    \end{tikzpicture}
    \caption{Logarithmic absolute error of the neural network and the linear volume fraction to the exact solution evaluated on a star ($R_s=8, A_s=5, n_s=4$) for different numbers
    of active symmetries $S$. (a) linear, $MAE = 2.157e-2$ (b) $S=0$, $MAE=1.322e-3$, (c) $S=4$, $MAE=6.957e-4$, (d) $S=8$, $MAE=6.349e-4$. The architecture of the used regression neural network is \textit{2 50}.
    The color scheme is ranging from maximum (blue) to minimum (red) absolute error.}
    \label{star_symmetry}
\end{figure}

As described previously, we preserve symmetry by mirroring and rotating the input levelset grid and subsequently averaging the predictions (compare Figure \ref{symmetries_fig}).
In Figure \ref{star_symmetry}, the logarithmic absolute error of the linear and neural network volume fraction to the exact values for a star shape is depicted. 
We compare different numbers of active symmetries. 
While the linear reconstruction is inherently symmetric, it can be seen that the error pattern goes from random ($S=0$) to fully
symmetric ($S=8$) for the neural network reconstruction. The corresponding mean absolute errors $MAE$ along the perimeter of the star are displayed in the caption of the figure.

\begin{figure}[!b]
    \centering
\begin{tikzpicture}

    \definecolor{color0}{rgb}{0.933333333333333,0.509803921568627,0.933333333333333}
    
    \begin{groupplot}[group style={group size=2 by 1, horizontal sep = 1cm}, width=5cm, height=4cm, scale only axis]
        \nextgroupplot[
            legend cell align={left},
            legend style={draw opacity=1, text opacity=1, nodes={scale=0.9}},
            log basis x={10},
            log basis y={10},
            tick align=outside,
            tick pos=left,
            title={volume fraction},
            xlabel={$R$},
            xmin=2.72270130807791, xmax=11614.4861384034,
            xmode=log,
            xtick style={color=black},
            ylabel={$MAE$},
            ymin=1e-05, ymax=0.1,
            ymode=log,
            ytick style={color=black}    ]  
    \addplot [semithick, red, mark=*, mark size=1, mark options={solid}]
    table {%
    3.98107170553497 0.00216056164053093
    7.94328234724281 0.000864217816180188
    15.8489319246111 0.000548840654360135
    31.6227766016838 0.000427755848083085
    63.0957344480193 0.000371121517743177
    125.892541179417 0.000354104537445238
    251.188643150958 0.000337710891050293
    501.187233627272 0.000340852140699119
    999.999999999999 0.000342156133568687
    1995.26231496888 0.000347243519446305
    3981.07170553497 0.000345909657327696
    7943.28234724281 0.000344079696030229
    };
    \addlegendentry{0}
    \addplot [semithick, blue, mark=*, mark size=1, mark options={solid}]
    table {%
    3.98107170553497 0.00186432562465115
    7.94328234724281 0.000658155019196056
    15.8489319246111 0.000411541348937333
    31.6227766016838 0.000320494320556873
    63.0957344480193 0.000273938502966934
    125.892541179417 0.000264304106115585
    251.188643150958 0.00025656374850004
    501.187233627272 0.000258624032746791
    999.999999999999 0.000259533789779451
    1995.26231496888 0.000260324664755402
    3981.07170553497 0.000258124902435988
    7943.28234724281 0.000259698729458739
    };
    \addlegendentry{2}
    \addplot [semithick, green!50.1960784313725!black, mark=*, mark size=1, mark options={solid}]
    table {%
    3.98107170553497 0.00163804140774317
    7.94328234724281 0.000524115000112661
    15.8489319246111 0.000329109705474677
    31.6227766016838 0.000242601884161436
    63.0957344480193 0.000215529582764025
    125.892541179417 0.000204533307629621
    251.188643150958 0.000200774317836168
    501.187233627272 0.000199001896797897
    999.999999999999 0.000200181071479116
    1995.26231496888 0.000203725686641767
    3981.07170553497 0.000203835351180245
    7943.28234724281 0.000200154433069857
    };
    \addlegendentry{4}
    \addplot [semithick, color0, mark=*, mark size=1, mark options={solid}]
    table {%
    3.98107170553497 0.00153436741987822
    7.94328234724281 0.000450982609189652
    15.8489319246111 0.000271620730758837
    31.6227766016838 0.000194136223689856
    63.0957344480193 0.000174035667988569
    125.892541179417 0.000170901294279878
    251.188643150958 0.000164756063442691
    501.187233627272 0.000164768503286403
    999.999999999999 0.000164862601365055
    1995.26231496888 0.000168421337695184
    3981.07170553497 0.000167957864830881
    7943.28234724281 0.0001646572034656
    };
    \addlegendentry{8}
    \addplot [semithick, black, mark=x, mark size=2, mark options={solid}]
    table {%
    3.98107170553497 0.0333463259006919
    7.94328234724281 0.0169472164652127
    15.8489319246111 0.0076863736641285
    31.6227766016838 0.00403074920292307
    63.0957344480193 0.00196799912121385
    125.892541179417 0.00102099085088734
    251.188643150958 0.000516046381622794
    501.187233627272 0.000256012754532068
    999.999999999999 0.000126727544951096
    1995.26231496888 6.41316410488598e-05
    3981.07170553497 3.08498914643931e-05
    7943.28234724281 1.59317106938867e-05
    };
    \addlegendentry{linear}
    
    \nextgroupplot[
        legend cell align={left},
        log basis x={10},
        log basis y={10},
        tick align=outside,
        tick pos=left,
        title={apertures},
        xlabel={$R$},
        xmin=2.72270130807791, xmax=11614.4861384034,
        xmode=log,
        ymin=1e-05, ymax=0.1,
        ymode=log,
        ytick style={color=black},
        yticklabels={,,},]  ]
    \addplot [semithick, red, mark=*, mark size=1, mark options={solid}]
    table {%
    3.98107170553497 0.00785579654364242
    7.94328234724281 0.00341071343602436
    15.8489319246111 0.00219690581485414
    31.6227766016838 0.00177802121762071
    63.0957344480193 0.00151123864313074
    125.892541179417 0.00147226392827916
    251.188643150958 0.00138520106534903
    501.187233627272 0.00136284083169003
    999.999999999999 0.00139703049467902
    1995.26231496888 0.00139576004488304
    3981.07170553497 0.00141042518110695
    7943.28234724281 0.00139161432939061
    };
    \addplot [semithick, blue, mark=*, mark size=1, mark options={solid}]
    table {%
    3.98107170553497 0.00654234036085304
    7.94328234724281 0.00256442908600829
    15.8489319246111 0.00164147539796665
    31.6227766016838 0.00130596481917929
    63.0957344480193 0.00111405517098208
    125.892541179417 0.0010561864900074
    251.188643150958 0.00101354110865341
    501.187233627272 0.000991754824334424
    999.999999999999 0.0010151823447132
    1995.26231496888 0.00101281211187383
    3981.07170553497 0.00104945627182935
    7943.28234724281 0.00102386677771284
    };
    \addplot [semithick, green!50.1960784313725!black, mark=*, mark size=1, mark options={solid}]
    table {%
    3.98107170553497 0.0057249924094694
    7.94328234724281 0.00206453570369139
    15.8489319246111 0.00130887204830505
    31.6227766016838 0.000980136069422082
    63.0957344480193 0.000827696596136489
    125.892541179417 0.000772310887287571
    251.188643150958 0.00074188219441617
    501.187233627272 0.000722454863987609
    999.999999999999 0.000759780807197398
    1995.26231496888 0.000753389486688219
    3981.07170553497 0.000777906366493836
    7943.28234724281 0.000765052881609097
    };
    \addplot [semithick, color0, mark=*, mark size=1, mark options={solid}]
    table {%
    3.98107170553497 0.00527920398665995
    7.94328234724281 0.00184589410686227
    15.8489319246111 0.00106438431715321
    31.6227766016838 0.000771759988772405
    63.0957344480193 0.000641264291030961
    125.892541179417 0.000600104173004235
    251.188643150958 0.00058658087320099
    501.187233627272 0.000550690151346756
    999.999999999999 0.000583644514971861
    1995.26231496888 0.000587405934214967
    3981.07170553497 0.000609907205944572
    7943.28234724281 0.000590379781922392
    };
    \addplot [semithick, black, mark=x, mark size=2, mark options={solid}]
    table {%
    3.98107170553497 0.0344950657056197
    7.94328234724281 0.0157957212202473
    15.8489319246111 0.00754406077190478
    31.6227766016838 0.00399895891302192
    63.0957344480193 0.00185839938359031
    125.892541179417 0.00095064876331666
    251.188643150958 0.000483206520569169
    501.187233627272 0.000237472234543528
    999.999999999999 0.0001195798216627
    1995.26231496888 6.0249815149818e-05
    3981.07170553497 2.99032014861816e-05
    7943.28234724281 1.50002865937656e-05
    };
    \end{groupplot}
    
    \end{tikzpicture}
    
\caption{Mean absolute error $MAE$ of pure neural network and linear interface reconstruction on the convergence dataset for different numbers of active symmetries. The radius $R$ represents the normalized (with cell size) radius of the interface shapes, namely circles and ellipses. For a detailed description of the 
    dataset, the reader is referred to section \ref{data generation}. The used architectures for the regression neural networks is \textit{2 50}.}
\label{symmetry_convergence}
\end{figure}
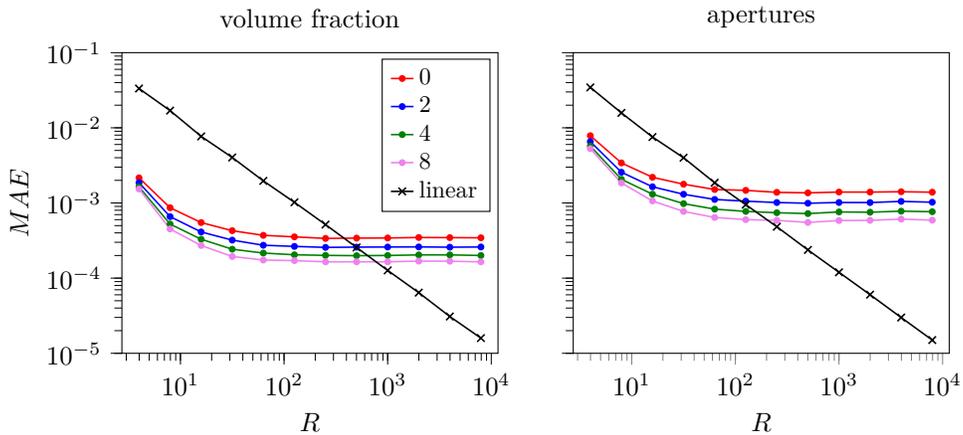

In Figure \ref{symmetry_convergence}, the mean absolute error of the neural network and linear reconstruction to the exact solution over the interface resolution $R$ for different numbers of active symmetries is displayed. Note that 
the radius represents a normalized (with cell size) value. The neural network reconstruction achieves accuracy improvements up to an order of magnitude for coarsely resolved
structures. This particular network (architecture \textit{2 50}) outperforms the linear reconstruction even for resolutions up to $R=500$ (volume fraction) and $R=200$ (apertures). 
It can be seen that the mean absolute error decreases successively with the number of active symmetries. For a detailed description of the used dataset for the evaluation of the convergence, the reader is referred to section \ref{data generation}.

\subsection{Network architectures and types}
\begin{figure}[!t]
    \centering
    \input{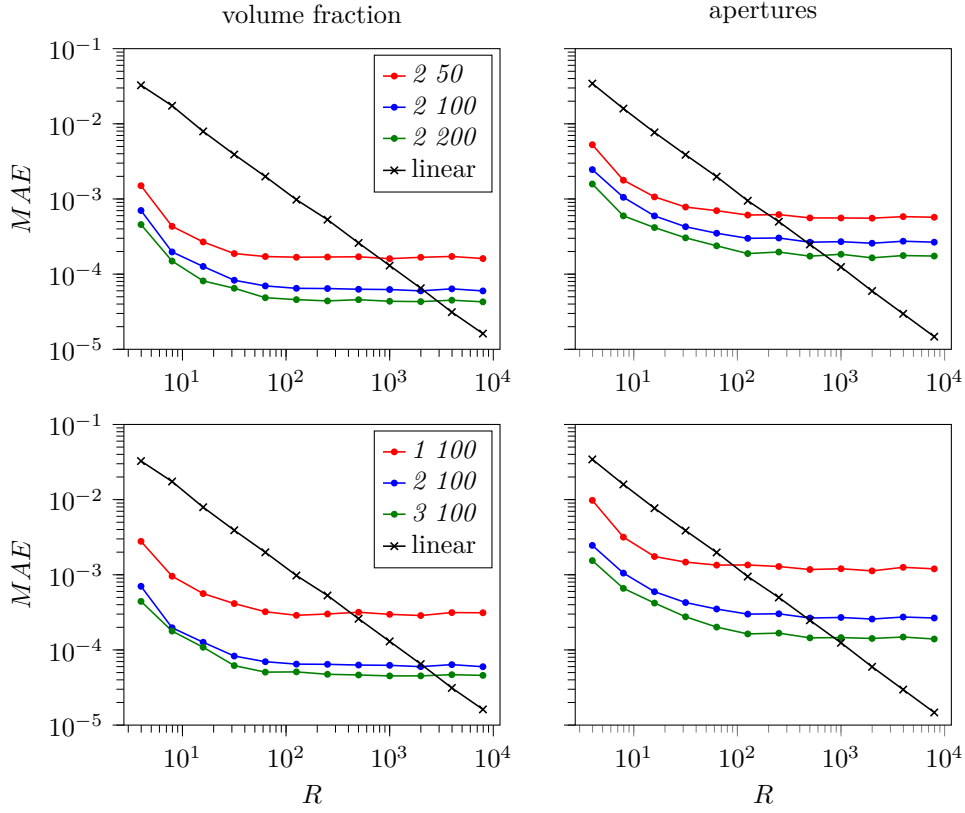}
\caption{Mean absolute error $MAE$ of pure neural network and linear interface reconstruction on the convergence dataset for different MLP architectures with $S=8$ active symmetries.
(For interpretation of the model architectures see section \ref{section:combined_model}.)}
\label{variation_convergence}
\end{figure}

In Figure \ref{variation_convergence}, the mean absolute error of the neural network and linear reconstruction to the exact solution over the interface resolution $R$ for various MLP architectures is displayed.
For this, we set the number of active symmetries to $S=8$. It is shown that the accuracy of the MLP is successively increasing with model complexity, i.e. with the number of hidden layers 
and with the number of nodes per layer. However, looking at the volume fraction in particular, the accuracy does not increase beyond a certain model complexity, indicating
that larger models than \textit{2 200} or \textit{3 100} will not lead to better performance, and merely increase cost.

In Table \ref{Training_validation_testing}, a comparison of the mean squared error $MSE$ on the training, validation and test set is displayed for the 
regression neural networks. Both MLP and CNN types for various architectures are shown. Additionally, the amount of parameters for each architecture is listed.
To provide a meaningful comparison between MLP and CNN, we only picked CNN architectures that result in similar amount of parameters compared to the MLP. The networks 
that are reasonably comparable across neural network types are highlighted with the same color.
It is observed that there is no significant difference in MSE between both neural network types. It may be necessary to increase the input levelset grid width for 
the CNN to be able to leverage its ability to account for the spatial structure.
Overall, the MLP shows slightly better performance with regard to training, validation and testing. 
Therefore, in the following, we will choose the MLP type for the regression task.


\setlength\arrayrulewidth{1pt}
\def\arraystretch{0.8}
\begin{table}[!t]
    \centering
    \adjustbox{max width=\textwidth}{
    \begin{tabular}{ c c  c  c  c  c}
        &&& \multicolumn{3}{c}{MSE} \\
        \cline{4-6} 
        Type & Architecture & Parameters & Training & Validation & Testing \\
        \hline
        \hline
        \multirow{6}{*}{MLP}
        & \textit{1 50 }& \color{blue!80} 551/704 & $5.08e-6/1.89e-4$ & $5.12e-6/1.92e-4$ & $3.34e-4/4.26e-3$ \\
        & \textit{1 100} & \color{violet!80} 1101/1404 & $3.16e-6/9.67e-5$ & $3.21e-6/9.92e-5$ & $2.85e-4/3.29e-3$ \\
        & \textit{1 200} & 2201/2804 & $1.44e-6/6.69e-5$ & $1.46e-6/6.96e-5$ & $2.36e-4/2.47e-3$ \\
        & \textit{2 50 }& \color{red!80} 3101/3354 & $1.00e-6/4.51e-5$ & $1.02e-6/4.69e-5$ & $1.96e-4/1.82e-3$ \\
        & \textit{3 50 }& 5651/5804 & $4.36e-7/1.98e-5$ & $4.57e-7/2.10e-5$ & $1.75e-4/1.11e-3$ \\
        & \textit{2 100} & \color{green!80} 11201/11504 & $2.27e-7/1.45e-5$ & $2.39e-7/1.57e-5$ & $1.64e-4/1.12e-3$ \\
        \hline
        \multirow{7}{*}{CNN}
        & \textit{3 16 1 20} & \color{blue!80} 521/584 & $5.42e-6/1.86e-4$ & $5.48e-6/1.89e-4$ & $3.93e-4/4.44e-3$ \\
        & \textit{3 8 1 50} & \color{blue!80} 581/734 & $5.55e-6/1.79e-4$ & $5.58e-6/1.81e-4$ & $4.10e-4/4.48e-3$ \\
        & \textit{3 32 1 20} & \color{violet!80} 1001/1064 & $3.47e-6/1.07e-4$ & $3.49e-6/1.09e-4$ & $3.08e-4/2.97e-3$ \\
        & \textit{3 16 1 50} & \color{violet!80} 1061/1214 & $2.90e-6/1.01e-4$ & $2.99e-6/1.03e-4$ & $2.75e-4/3.24e-3$ \\
        & \textit{3 8 1 100} & \color{violet!80} 1081/1384 & $3.16e-6/1.36e-4$ & $3.20e-6/1.38e-4$ & $2.84e-4/4.12e-3$ \\
        & \textit{3 8 2 50}& \color{red!80} 3131/3284 & $1.70e-6/9.40e-5$ & $1.71e-6/9.60e-5$ & $2.30e-4/3.18e-3$ \\
        & \textit{3 16 2 100} & \color{green!80} 12061/12364 & $3.18e-7/1.58e-5$ & $3.36e-6/1.68e-5$ & $1.77e-4/1.17e-3$ \\
    \end{tabular}
    }
    \caption{Mean squared error $MSE$ of various regression neural network models for volume fraction/aperture on the training, validation and test set. (For interpretation of the model architectures see section \ref{section:combined_model}.)}
    \label{Training_validation_testing}
\end{table}

\subsection{Combined model}
\begin{figure}[!b]
    \centering
    \input{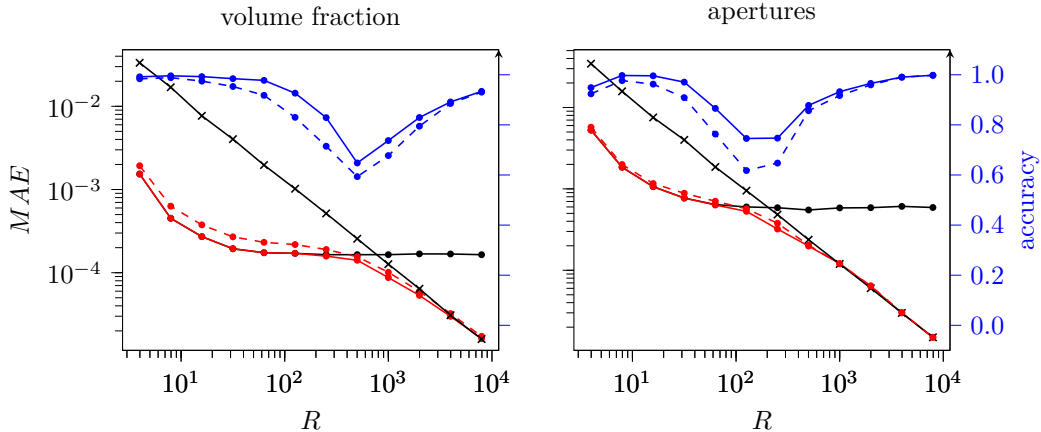}
\caption{Mean absolute error $MAE$ of the combined model and the linear interface reconstruction on the convergence dataset.
The black lines indicate the $MAE$ of the linear (cross marker) and the pure neural network (circle marker) reconstruction.
The red lines indicate the $MAE$ of the combined model. The blue lines represent the mean classifier accuracy. The dashed  and solid  lines distinguish between the classification neural network architectures \textit{3 2 1 5} and \textit{5 2 1 5}, respectively.
The used architecture for the regression neural networks is \textit{2 50}.}
\label{combined_model_convergence}
\end{figure}
As described in section \ref{section:combined_model}, the combined model consists of a classification neural network deciding whether to use a neural network or the linear approach to reconstruction the interface,
depending on its resolution. In Figure \ref{combined_model_convergence}, the mean absolute error of the combined model and the linear interface reconstruction to the exact solution over the interface resolution $R$ 
is depicted. Furthermore, the mean accuracy of the classifier is shown. We compare two classification neural network architectures. In particular, the architectures 
differ only in the kernel width, i.e. the input levelset grid. We observe a significant increase of classification accuracy for the $5\times5$ input grid, compared to the $3\times3$ input grid.
The combined model uses the neural network reconstruction for coarsely resolved structures and recovers the linear interface reconstruction upon mesh refinement.

\section{Model performance coupled with CFD solver} \label{model perfomance solver}

\subsection{Implementation details}
The feed-forward operations of the ML models mainly consist of basic linear arithmetic operations, i.e. matrix/matrix multiplications and matrix/vector additions.
For optimal performance, the neural networks are implemented using blaze, a high-perfomance C++ library for dense matrix arithmetic, making use of e.g. SIMD instructions, data structure padding, and smart cache usage.
In particular, we are using Intel MKL as BLAS backend.

Weights and biases are stored in dynamic containers and can thus be loaded at runtime.
This increases flexibility regarding the usage of varying model complexities without the need to recompile the executable.
However, the amount of active symmetries is set at compile time for optimal performance.

ALPACA \cite{Hoppe} is a block-based multiresolution solver. A block is the smallest computational unit with regard to contiguous memory. The computational domain
consists of multiple blocks and a single block is usually composed of 16 to 32 cells, referred to as internal cells. Thus, within the solver, there exist an outer loop over the blocks and an inner loop over the internal cells.
This implies that the computation of the volume fraction and the apertures happens cell-wise. MPI parallelization is realized by assigning ranks to multiple blocks. 
When using a neural network however, predicting single samples at a time is slow. To make optimal use of BLAS, matrix/matrix multiplications should be favored over (multiple) matrix/vector multiplications.
In other words, performing a single feed forward for multiple samples (cells) is faster than performing a feed forward for each sample (cell). To make use of this within ALPACA, there exist in principle two options: 
1) Gather all cut cell indices within the whole computational domain and subsequently do a single feed-forward operation for the whole domain. 
2) Gather all cut cells within each block and perform a feed-forward operation for each block.
While option 1) is faster when solely considering the speed of the neural network, this approach has a drawback: It introduces communication overhead between ranks as the rank performing the feed forward must receive all cut cell indices and subsequently
must send the result to the corresponding ranks. Furthermore, this implementation requires major code restructure. We therefore implement the first option.

\begin{figure}[!t]
    \centering
    \adjustbox{max width=\textwidth}{
    \begin{tikzpicture}
        \node at (0.0,0) {\includegraphics[scale=0.17]{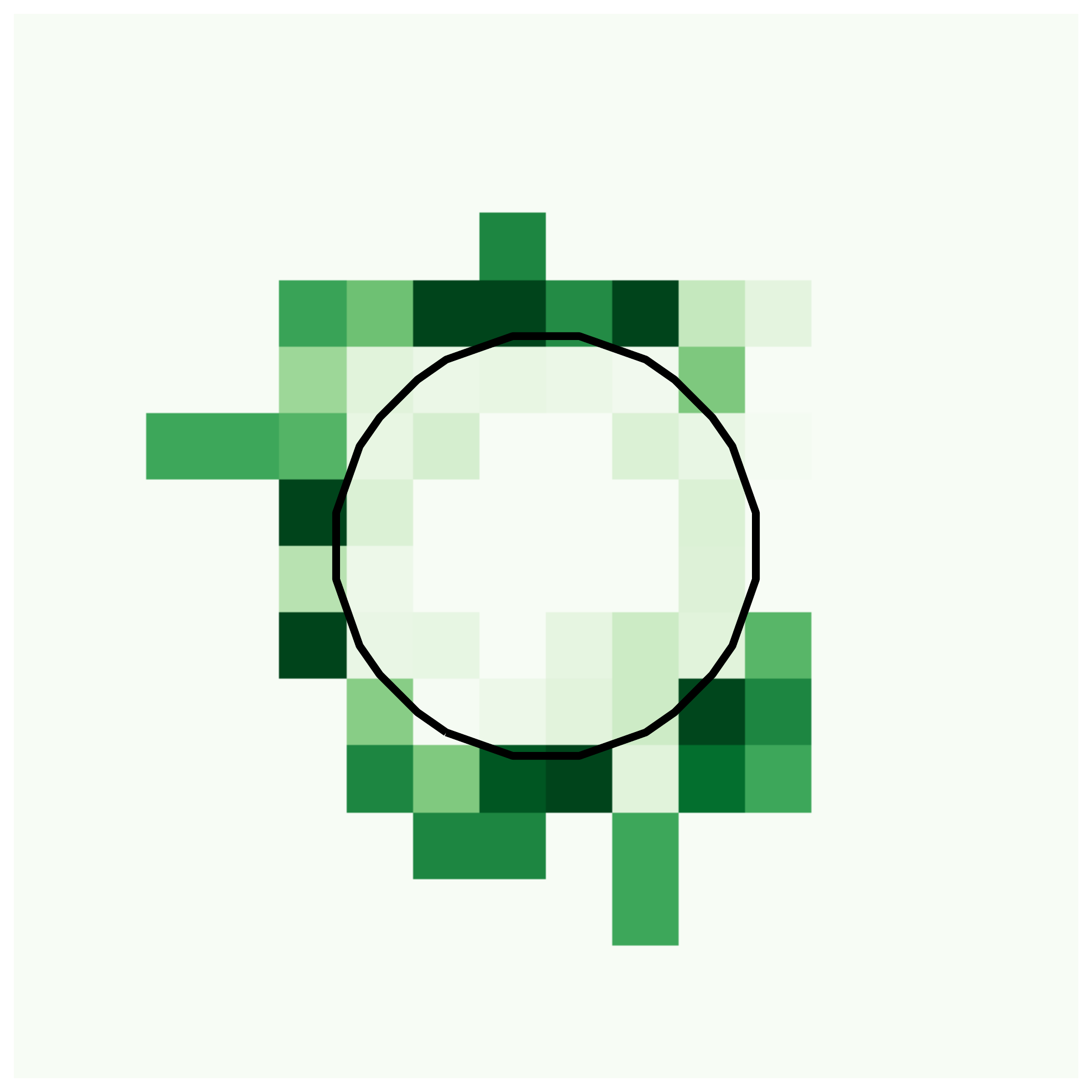}};
        \node at (0.0,2) {$S = 0$};
        \node at (4.0,0) {\includegraphics[scale=0.17]{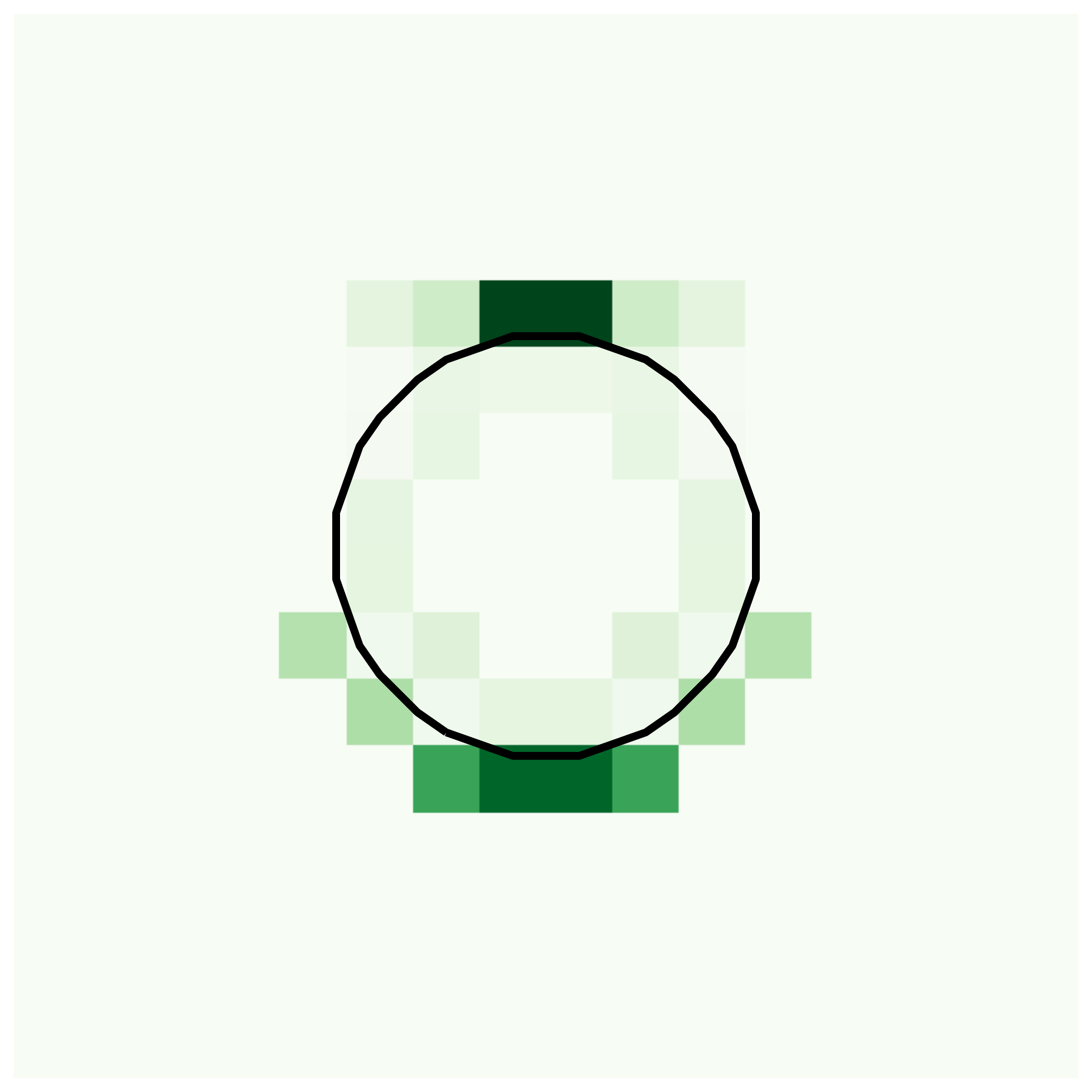}};
        \node at (4.0,2) {$S = 2$};
        \node at (8.0,0) {\includegraphics[scale=0.17]{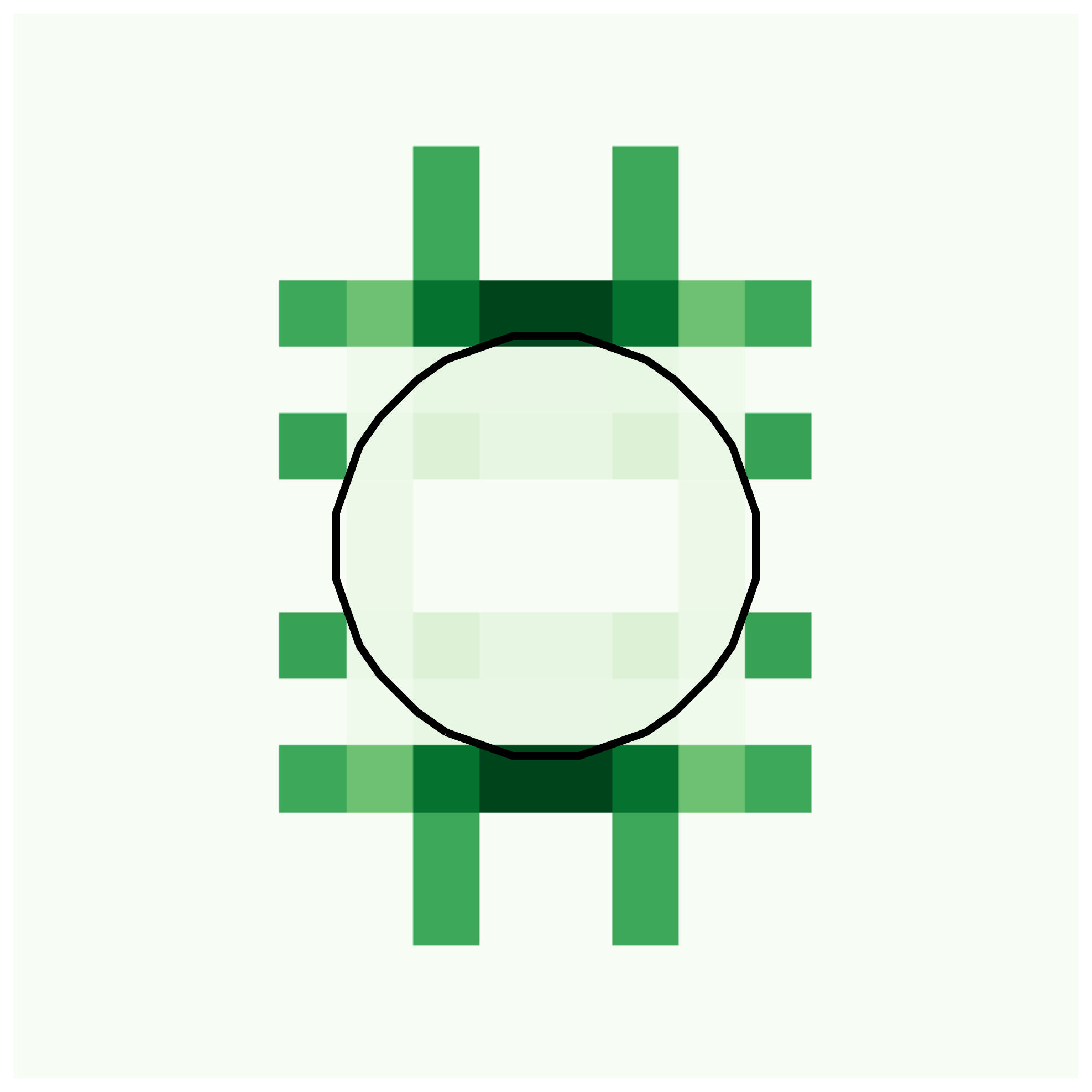}};
        \node at (8.0,2) {$S = 4$};
        \node at (12,0) {\includegraphics[scale=0.17]{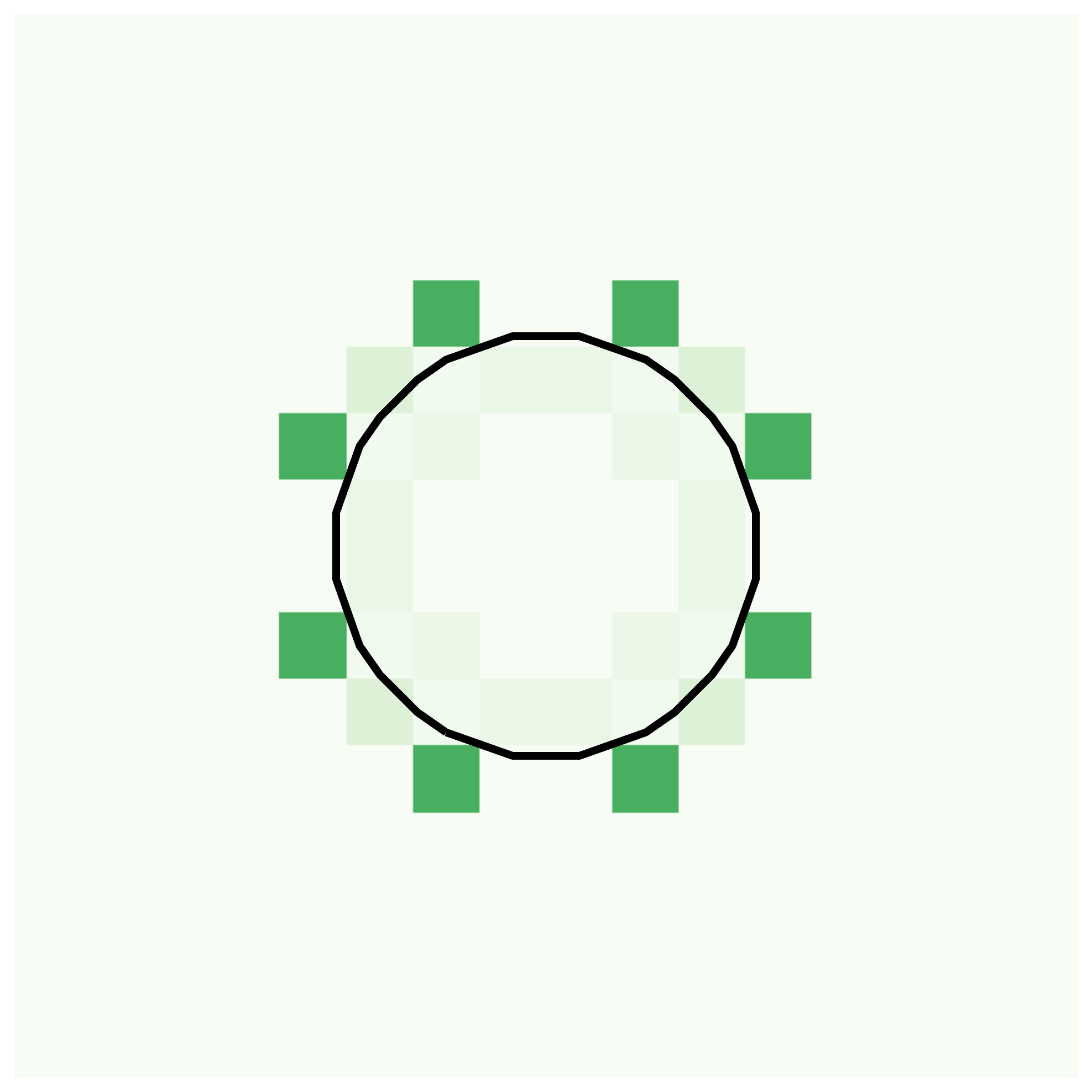}};
        \node at (12.0,2) {$S = 8$};
    \end{tikzpicture}
    }
    \caption{Interface noise (absolute velocity) due to floating point errors for a circular interface for varying amounts of active symmetries $S$.}
    \label{floating_point_symmetry}
\end{figure}
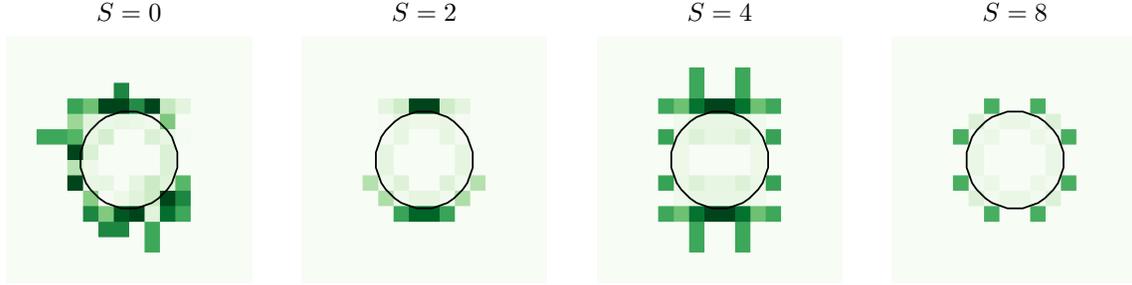

The apertures need special treatment with regard to symmetry preservation. This is due to the fact that the aperture predictions of a cell face shared by 
two adjacent cut cells are not equal for both input levelset grids. Hence, it is necessary to average the aperture predictions across cell faces.
This is done by simply computing the mean value of the aperture at a cell face shared by two adjacent cut cells.
Additionally, symmetry breaking floating point errors are addressed by sorting the symmetry predictions in descending order before computing the mean.

To show the floating point symmetric implementation, we simulate a static drop without surface tension in a single block of $16\times16$ cells. Floating point errors resulting from the computations at the interface are displayed in Figure \ref{floating_point_symmetry}.
In particular, the absolute velocity is illustrated. It can be seen that, depenending on the amount of active symmetries (compare Figure \ref{symmetries_fig}), the velocity noise shows a random ($S=0$) to fully symmetrical ($S=8$) pattern. 


\subsection{Efficieny comparison}

\begin{table}[!b]
    \centering
    \begin{tabular}{r | c c c c }
        & \multicolumn{4}{c}{Symmetries} \\
        \cline{2-5} 
        Architecture & 0 & 2 & 4 & 8  \\
        \hline
        \hline
        & \multicolumn{4}{c}{classifier inactive} \\
        \cline{2-5} 
        \textit{1 50 } & 0.39 / 1.00 & 0.97 / 1.17 & 1.35 / 1.97 & 2.66 / 3.69 \\
        \textit{1 100} & 0.84 / 0.89 & 0.82 / 1.26 & 1.65 / 2.29 & 3.31 / 5.54 \\
        \textit{2 50 } & 0.69 / 1.82 & 1.98 / 1.73 & 2.57 / 3.17 & 5.11 / 6.15 \\
        \textit{3 50 } & 1.00 / 2.02 & 3.24 / 2.43 & 3.83 / 4.42 & 7.55 / 8.61 \\
        \textit{2 100} & 2.50 / 1.69 & 2.25 / 2.65 & 4.38 / 5.11 & 8.99 / 12.00 \\
        \cline{2-5} 
        & \multicolumn{4}{c}{classifier active} \\
        \cline{2-5} 
        \textit{1 50 } & $4.89$ / $5.78$ & $7.62$ / $7.32$ & $9.78$ / $10.76$ & $18.3$ / $19.54$ \\
        \textit{1 100} & $5.75$ / $5.53$ & $6.62$ / $7.29$ & $10.79$ / $11.69$ & $19.92$ / $24.61$ \\
        \textit{2 50 } & $5.54$ / $7.07$ & $9.35$ / $8.73$ & $12.9$ / $12.9$ & $23.16$ / $24.24$ \\
        \textit{3 50 } & $6.54$ / $8.03$ & $11.51$ / $10.15$ & $15.25$ / $16.10$ & $28.63$ / $30.15$\\
        \textit{2 100} & $9.35$ / $8.00$ & $10.8$ / $11.7$ & $19.13$ / $19.17$ & $35.04$ / $42.11$ \\

    \end{tabular}
    \caption{Comparison of wall clock time to fill the volume fraction and aperture . The table shows the ratio $\sfrac{T_{nn}}{T_{linear}}$, where $T_{nn}$ and $T_{linear}$ are
    the wall clock times required by the neural network and the linear interface reconstruction, respectively. The used classifier architecture is \textit{5 2 1 5}. (For interpretation of the model architectures see section \ref{section:combined_model})}
    \label{wallclock_buffer}
\end{table}

We assess the speed of the proposed method by comparing the wall clock time $T$ necessary to fill the volume fraction and aperture buffer 
for a circular interface within a single block of $16\times 16$ cells (compare Figure \ref{symmetries_fig}). Of the 256 cells within the block, $14\%$ are cut cells, i.e. cells 
for which an interface reconstruction is performed. We average the wall clock time over 10000 runs on a Intel i5-4690K CPU.

In Table \ref{wallclock_buffer},
the ratio $T_{nn}/T_{linear}$ is depicted, with $T_{nn}$ and $T_{linear}$ being the wall clock time for the neural network and the linear interface reconstruction,
respectively. In particular, the values for the volume fraction / aperture neural network are listed seperately. It is distinguished between active and inactive classifier, i.e. combined and pure neural network reconstruction. Note that the filling of the aperture buffer 
in the neural network reconstruction requires additional overhead compared to the linear reconstruction due to the averaging procedure across cell faces.
Furthermore, while these values give insight into the sample-wise speed of the proposed approach, they are not representative for full simulations.
This is due to the fact that the wall clock time needed to perform the interface reconstruction is minor compared to the computational time for a single time step.


\subsection{Rising bubble}

\begin{figure}[!b]
    \centering
    \begin{tikzpicture}
    \coordinate (A) at (-4.8,0);
    \coordinate (C) at (3.5,7);
    \coordinate (X) at ($.5*(C)+(A)$);
    \coordinate (Y) at ($.25*(C)$);
    \coordinate (Y1) at ($.5*(C)$);
    \coordinate (X1) at ($(A)+(C)$);
    
    \draw[line width=2pt, fill=black!20!white] (A) rectangle ($(A) + (C)$);
    \draw[line width=1pt, fill=white] (X|-Y) circle (0.875);
    \node[yshift=0.5cm] at (X|-Y) {fluid 1};
    \node[above, xshift=0.8cm] at (A) {fluid 2};
    \draw[->] ($(X|-Y)+(0,4)$) -- ($(X|-Y)+(0,3.5)$) node[right, midway] {g};
    \dimline[extension start length=1cm, extension end length=1cm,extension style={black}] {($(A|-C)+(0,1)$)}{($(X1|-C)+(0,1)$)}{1.0};
    \dimline[extension start length=1cm, extension end length=1cm,extension style={black}, label style={rotate=-90}] {($(A)+(-1,0)$)}{($(A|-C) + (-1,0)$)}{2.0};
    \dimline[extension start length=-1.3cm, extension end length=-1.3cm,label style={rotate=-90,fill=black!20!white}, extension style={black}] {($(X|-Y)-(0,0.875)+(1.3,0)$)}{($(X|-Y)+(0,0.875)+(1.3,0)$)}{0.5};
    \dimline[extension start length=0, extension end length=0, label style={rotate=-90, fill=black!20!white, near start}] {(X|-A)}{(X|-Y)}{0.5};
    \dimline[extension start length=0, extension end length=0, label style={fill=black!20!white, xshift=-0.4cm}] {(Y-|A)}{(X|-Y)}{0.5};
    \node[rotate=90, above] at (A|-Y1) {periodic};
    \node[rotate=90, below] at (X1|-Y1) {periodic};
    \node[above] at (X|-C) {$u=0$, $v=0$, $p_\infty$}; 
    \node[below] at (X|-A) {$u=0$, $v=0$}; 
    \draw[->, line width = 1pt] (X) -- ($(X) + (0.6,0)$) node[at end, below] {$x$};
    \draw[->, line width = 1pt] (X) -- ($(X) + (0,0.6)$) node[at end, left] {$y$};
\end{tikzpicture}
    \caption{Schematic of the rising bubble setup.}
    \label{risingbubble_setup}
\end{figure}
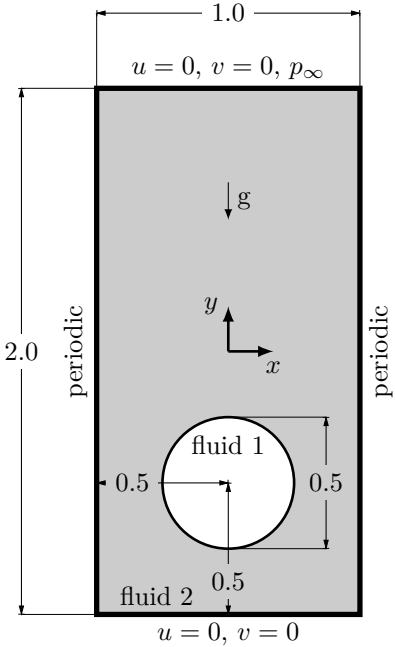

\begin{figure}[!t]
    \centering
    \adjustbox{max width=\textwidth}{
    \begin{tikzpicture}
        \node at (0.0,0) {\includegraphics[scale=0.10]{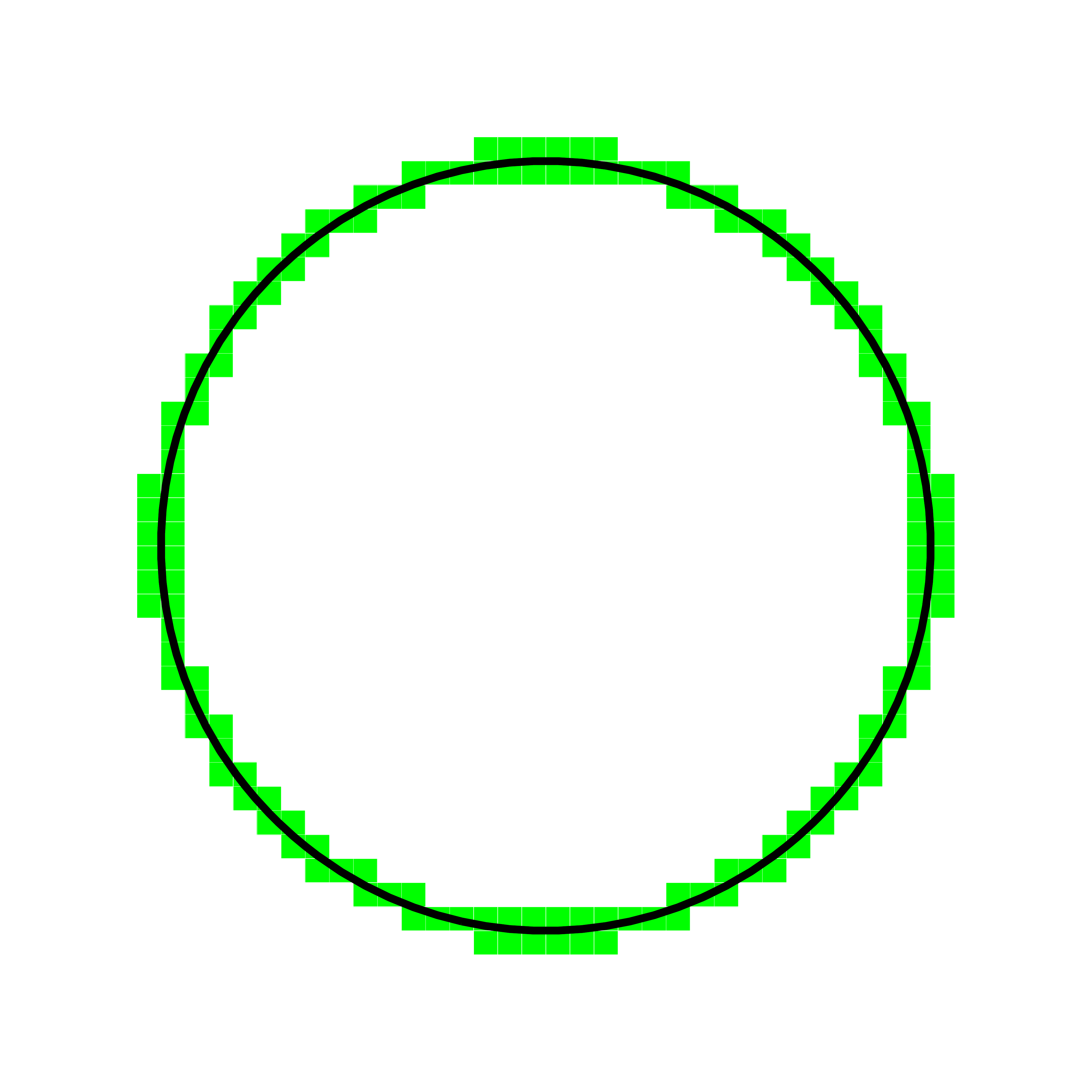}};
        \node at (3.5,0) {\includegraphics[scale=0.10]{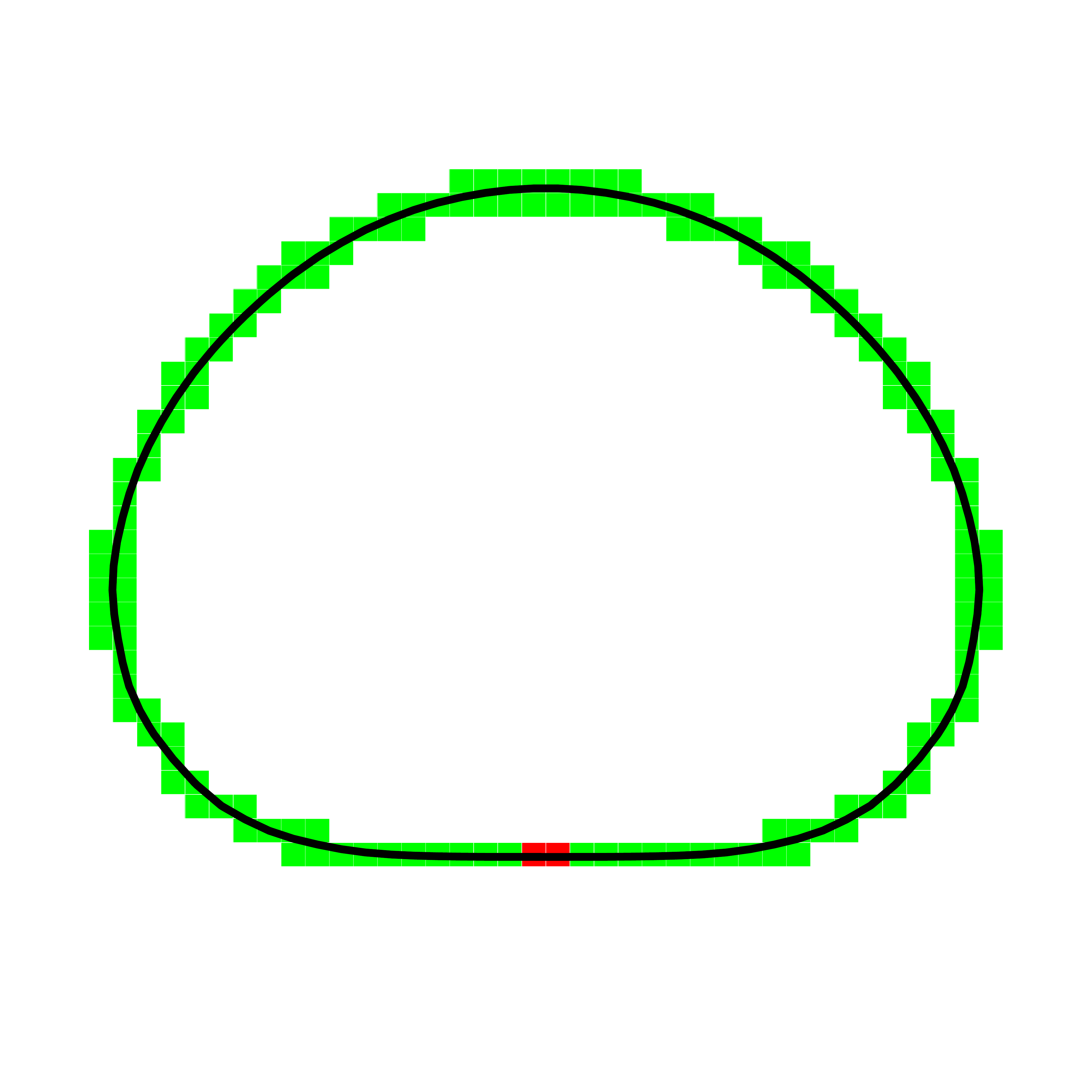}};
        \node at (7.5,0) {\includegraphics[scale=0.10]{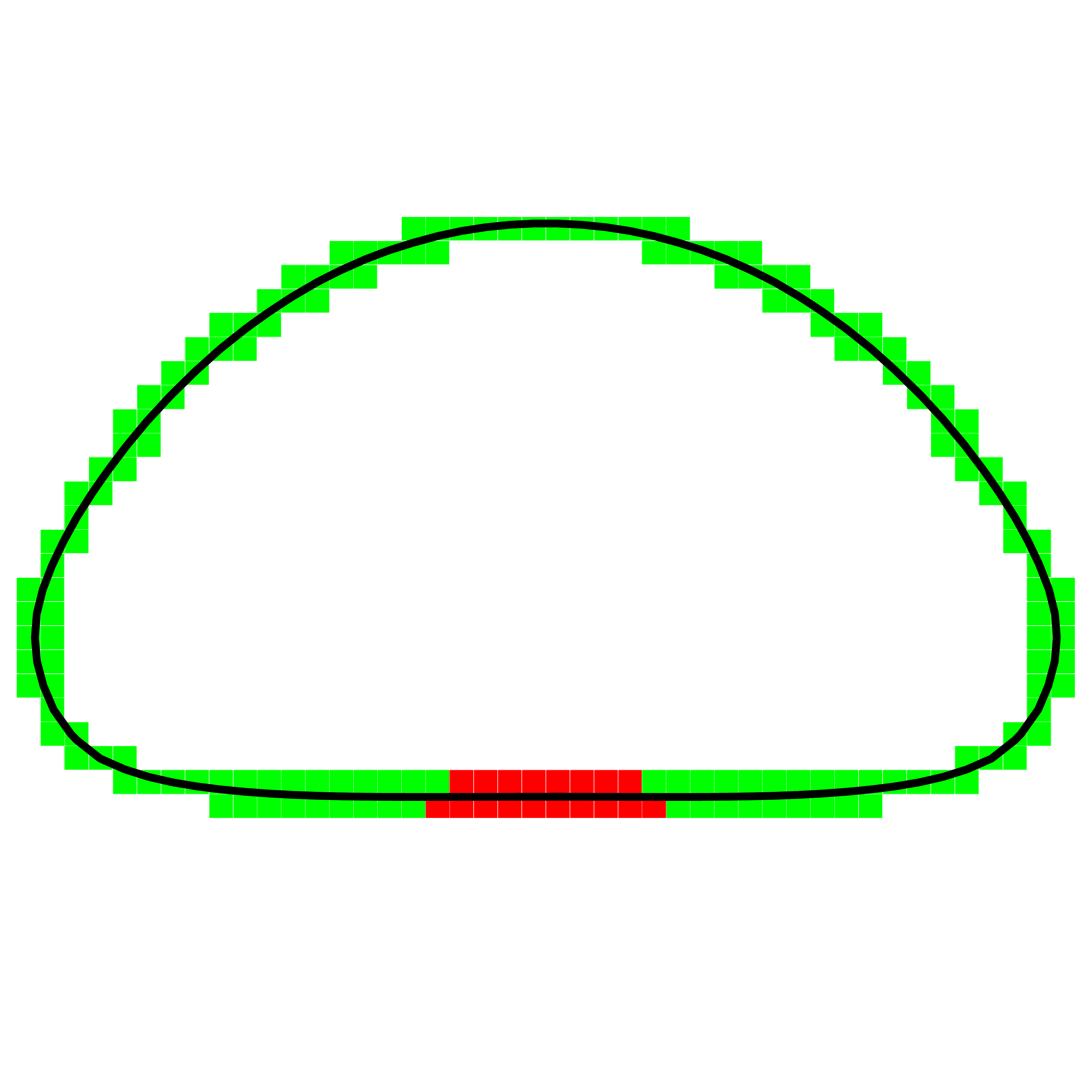}};
        \node at (11.7,0) {\includegraphics[scale=0.10]{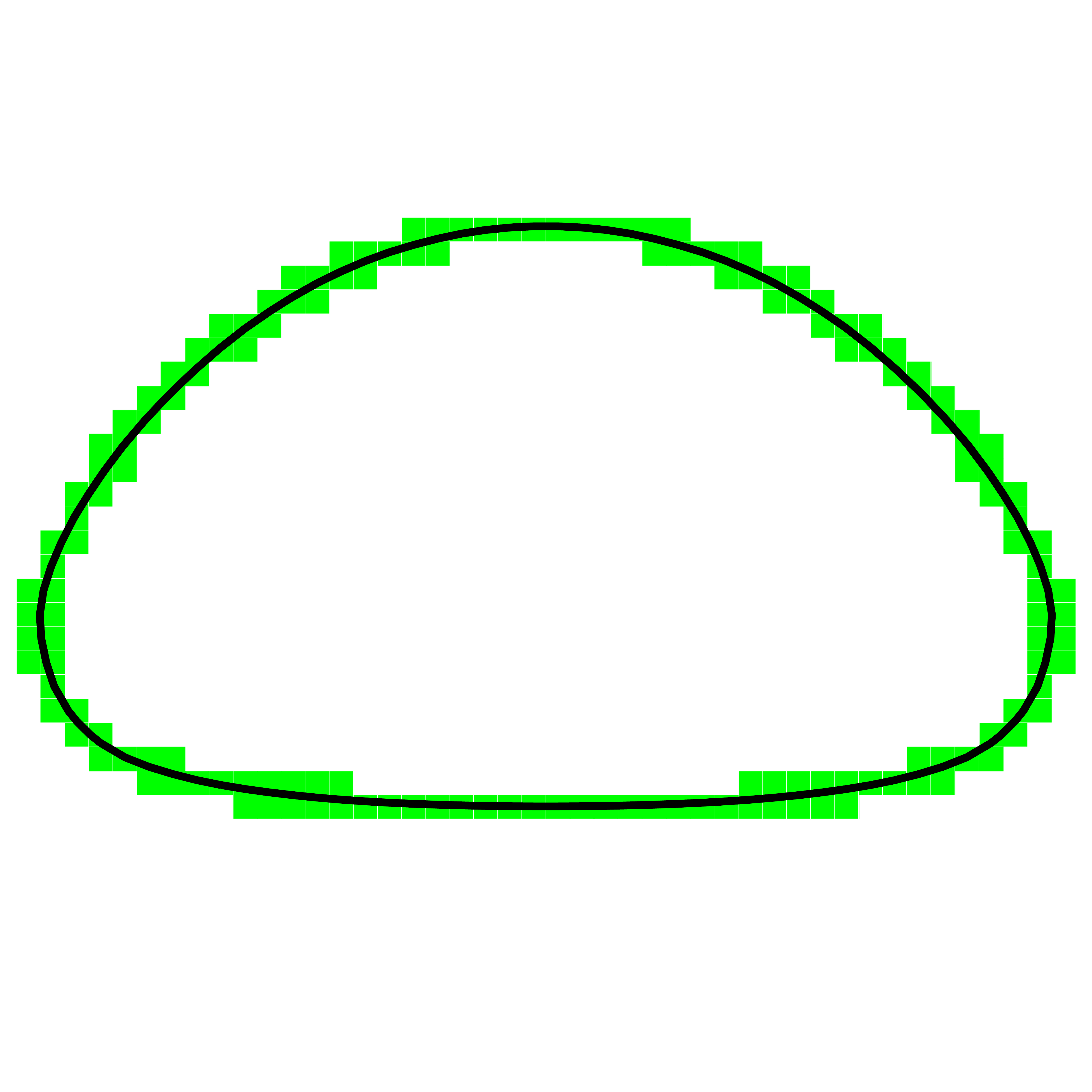}};
        \node at (0.0,-4) {\includegraphics[scale=0.10]{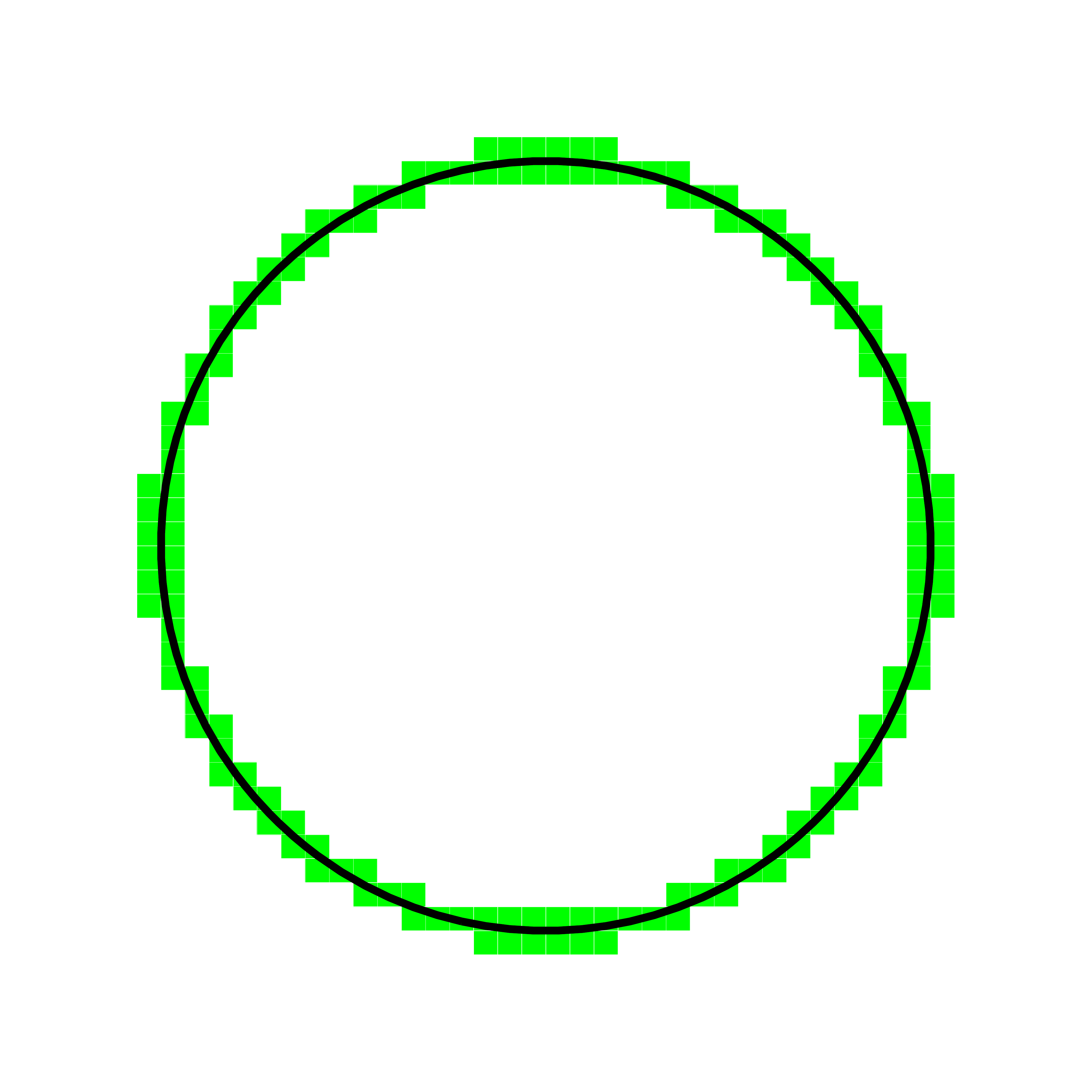}};
        \node at (3.5,-4) {\includegraphics[scale=0.10]{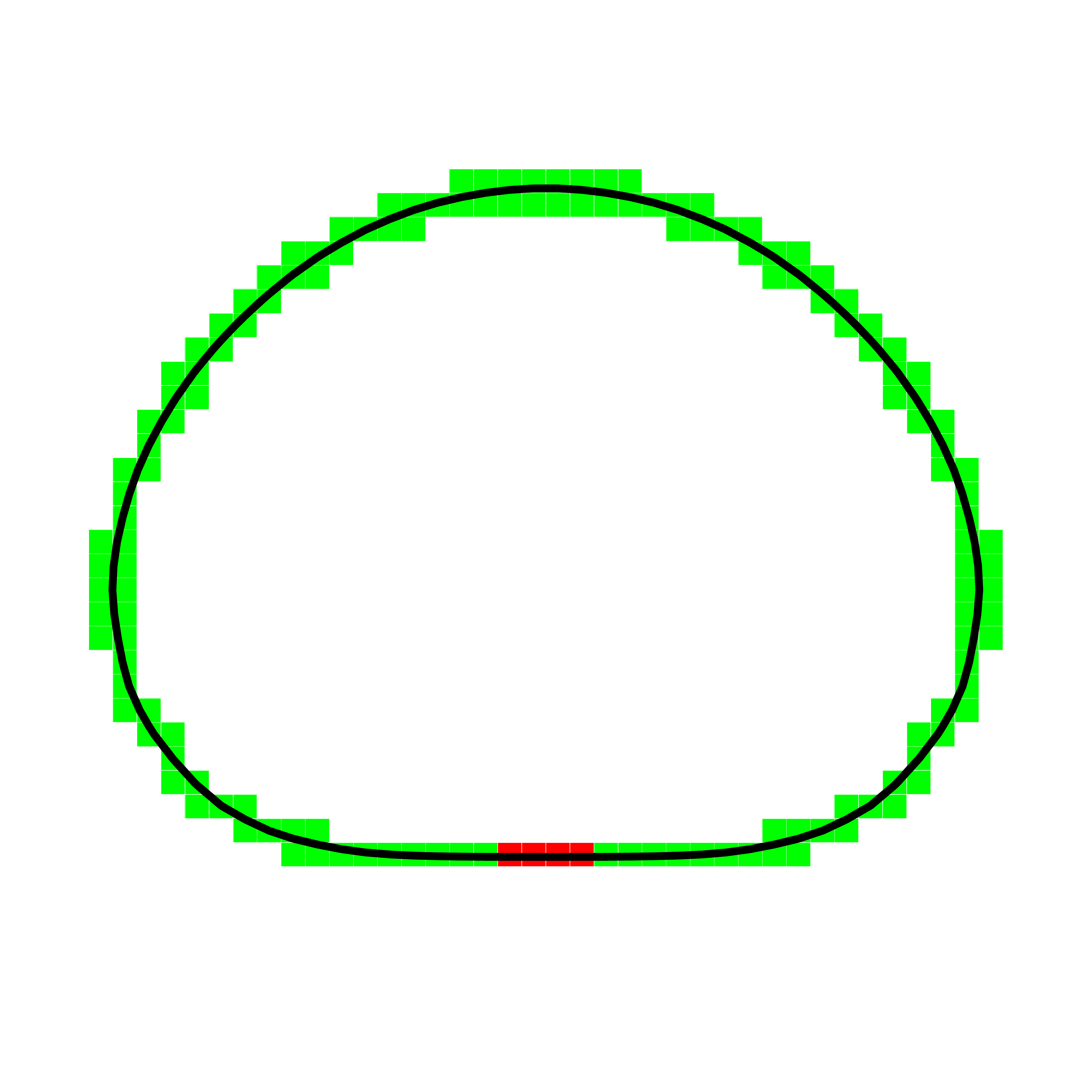}};
        \node at (7.5,-4) {\includegraphics[scale=0.10]{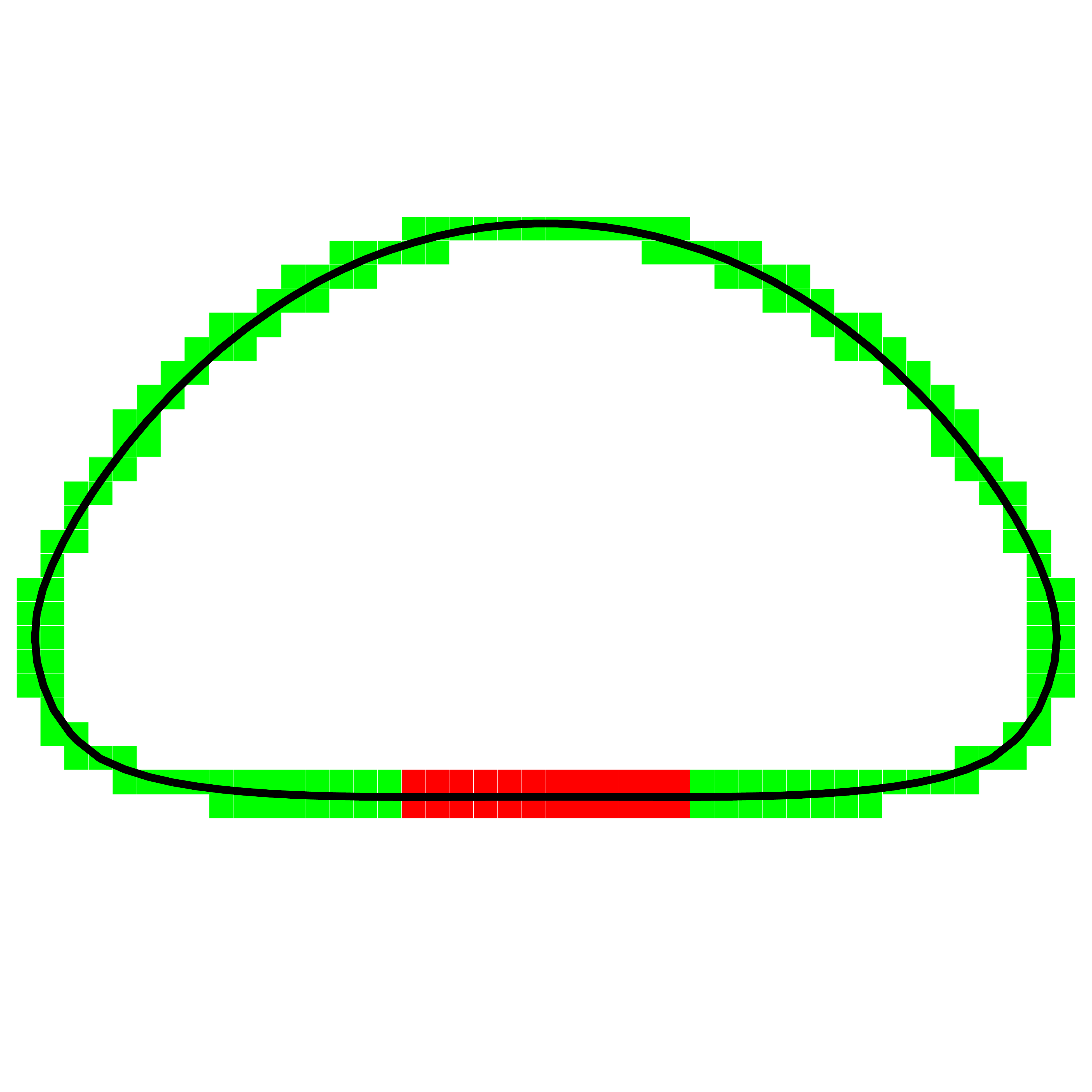}};
        \node at (11.7,-4) {\includegraphics[scale=0.10]{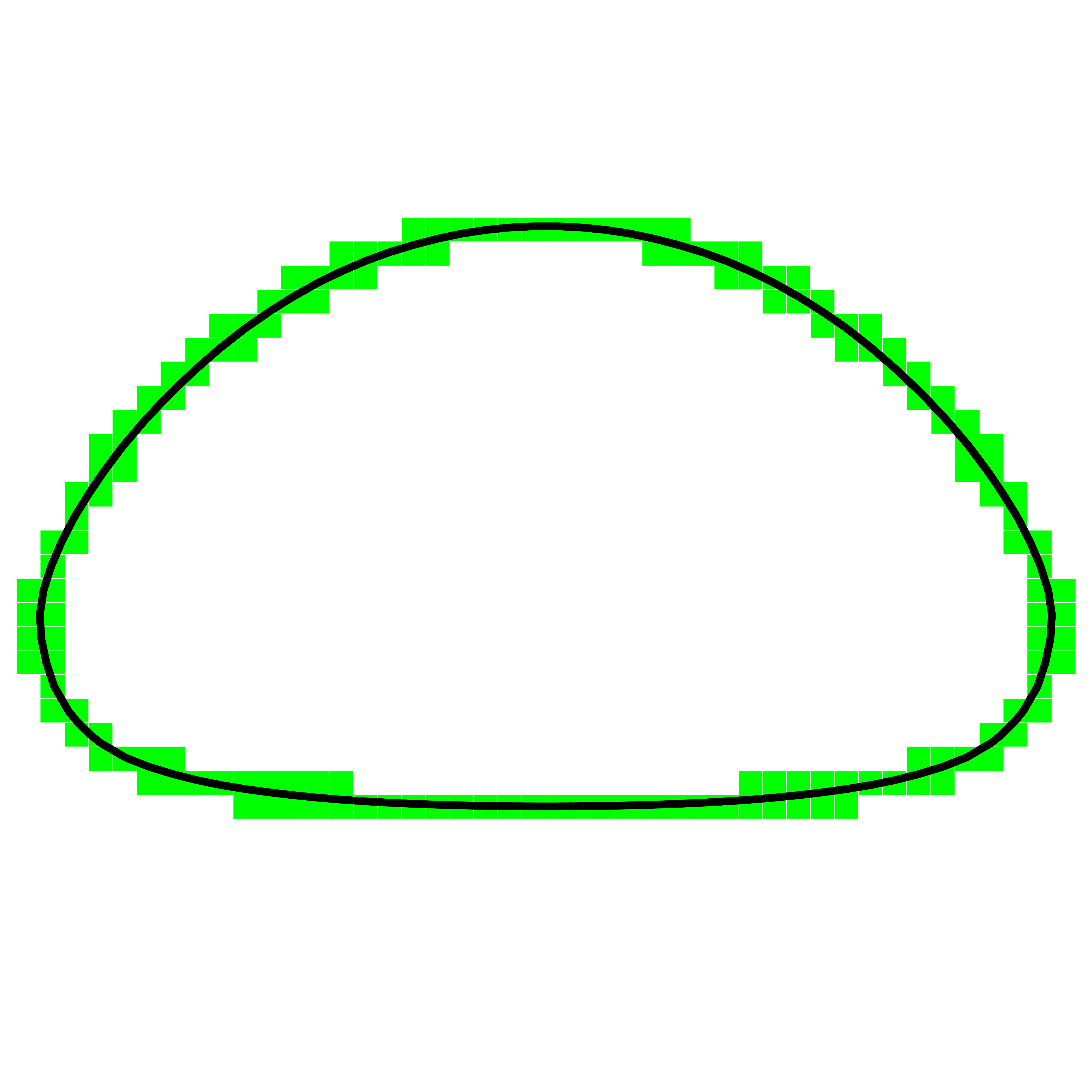}};
        \draw[->, line width=1pt] (5.3,2.3) -- (7.3,2.3) node[above, midway] {time};
        \node at (-1.9,0) [rotate=90] {volume fraction};
        \node at (-1.9,-4) [rotate=90] {apertures};
    \end{tikzpicture}
    }
    \caption{Rising bubble at times $t\in\{0.0,1.0,2.5,3\}$. The color of the cells indicate the decision of the classification neural network.
    In particular, green cells indicate neural network interface reconstruction and red cells indicate linear interface reconstruction.}
    \label{rising_bubble_classifier}
\end{figure}
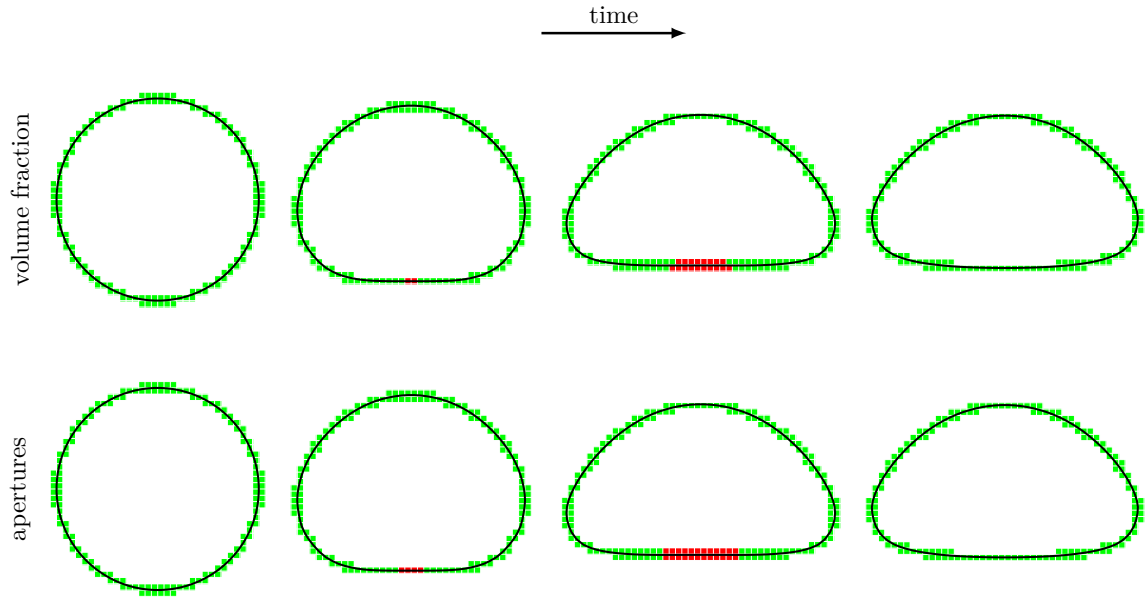

\begin{figure}[!t]
    \centering
    \input{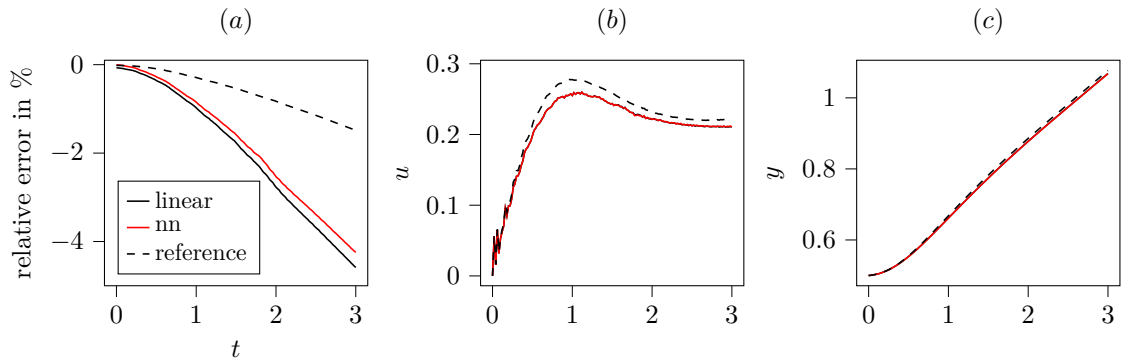}
    \caption{Comparison of mass, rise velocity and center of mass of the rising bubble (from left to right) for neural network and linear interface reconstruction. 
    The reference result is generated with the linear reconstruction on a mesh with twice the resolution.}
    \label{rising_bubble_stats}
\end{figure}

A bubble is immersed into another, heavier fluid. Buoyancy causes the bubble to rise and subsequently undergo moderate shape deformations. This test case 
was studied by \cite{Hysing2009b} as a benchmark case to compare various two-phase solver. Figure \ref{risingbubble_setup} shows a schematic of the initial 
and boundary conditions on the left. The two fluids are bound by the spatio-temporal domain $(x,y)\in[-0.5,0.5]\times[-1.0,1.0]$ and $t\in[0.0,3.0]$, with 
a resolution of $64\times128$. The initial bubble radius is $R=0.25$. 
The densities and viscosities are $\rho_1=100$, $\rho_2=1000$, $\mu_1=1$ and $\mu_2=10$, respectively. The surface tension coefficient is $\sigma=24.5$ and the gravitational acceleration 
is $\mathbf{g}=[0,-0.98]^T$. At the north and south boundary, the no slip condition $u=0$ and $v=0$ is imposed, while the east and west boundaries are periodic.
To simulate this case, we choose the regression neural network \textit{2 50} and the classifier \textit{5 2 1 5} with active symmetries $S=8$. 

In Figure \ref{rising_bubble_classifier}, the temporal evolution of the interface is depicted. We highlight the cut cells, where the classifier 
chooses a neural network/linear reconstruction in green/red. In Figure \ref{rising_bubble_stats}, a comparison of the exact mass, the rising velocity and 
the center of mass over time is illustrated. The combined model runs stably and achieves more accurate results with regard to the mass.

In Table \ref{rising_bubble_speed}, we compare the wall clock time needed to perform the simulate the rising bubble. Note, that this simulation was run on two CPUs.
For larger simulations, the overhead resulting from increased communication between hundres of CPUs would result in a significant reduction of the reported
values.








\begin{table}
    \centering
    \begin{tabular}{r | c c c c }
        & \multicolumn{4}{c}{Symmetries} \\
        \cline{2-5} 
        Architecture & 0 & 2 & 4 & 8  \\
        \hline
        \hline
        & \multicolumn{4}{c}{classifier inactive} \\
        \cline{2-5} 
        \textit{2 50} & $+2.1\%$& $+6.4\%$ & $+7.7\%$ & $+13.8\%$ \\
        \cline{2-5} 
        & \multicolumn{4}{c}{classifier active} \\
        \cline{2-5} 
        \textit{2 50} &$+11.44$\%& $+16.0\%$ & $+23.6\%$ & $+38.6\%$ \\ 
    \end{tabular}
    \caption{Comparison of wall clock time for the rising bubble for different amount of active symmetries.
    The table shows the relative difference in percent to the linear interface reconstruction. The used classifier architecture is \textit{5 2 1 5} (For interpretation of the model architectures see section \ref{section:combined_model}.)}
    \label{rising_bubble_speed}
\end{table}


\section{Conclusion}\label{conclusion}

Machine learning (ML) has recently been used to improve certain parts of conventional CFD codes, e.g. the cell-face reconstruction in finite-volume solvers or the curvature computation in the 
Volume-of-Fluid method. 
While the latter improved the accuracy for coarsely resolved interfaces, the resulting ML models do not provide convergence or symmetry. 
In this work, we shift the focus on the interface reconstruction in the Levelset method, i.e. the computation of the volume fraction and the portions of the fluids along the cell faces, 
and adress aforementioned shortcomings. 
We propose a combined model, consisting of a classification neural network that chooses between conventional (linear) and neural network reconstruction. 
While the linear interface reconstruction is inherently symmetric and has first order convergence, it yields poor results for coarsely resolved interfaces. 
Our experiments show that the pure neural network reconstruction delivers much better approximations in the coarse regime.
However, the error level of the neural network model does not converge upon mesh refinement.
The classification neural network decides, based on the given levelset field, whether evaluation of the conventional or the neural network model is preferable.
For coarsely resolved interfaces the classifier chooses the neural network reconstruction and switches to the conventional model for finely resolved structures.
Thereby, the proposed combined approach achieves signficant accuracy improvements for resolutions up to 200 cells per effective radius of the interface, 
while at the same time recovering the first order convergence of the linear interface method in the asymptotic limit.
Symmetries are preserved by evaluating the neural network on mirrored and rotated variations of the original levelset field and subsequently averaging over these predictions.
The combined model is implemented into ALPACA, a block-based multiresolution solver for compressible flows.
For the rising bubble test case, the proposed model is stable and preserves the mass of the bubble better than the conventional model.
The perfomance in the solver depends on the depth of the neural network and the number of active symmetries.

In multiphase flows with low Weber numbers the surface tension plays a significant role.
Initial results \ref{appendix_curvature} of curvature neural networks showed that these models yield better results than the conventional method for circular interfaces in the coarse regime.
However, to make the use of neural networks for the curvature feasable, higher accuracy improvements have to be achieved for a broader range of interface shapes. 
This is left for future work.
Validation of the proposed data-driven model with more complex test cases, e.g. test cases with merging interfaces, and the extension to 3D simulations are subject of current research.

\section*{Acknowledgements}
This project has received funding from the European Research Council (ERC) under the European Union's Horizon 2020 research and innovation programme (grant agreement No. 667483).

\appendix

\section{Model training}
\label{sec:ModelTraining}

In the following, we describe details of the neural network training process.
All networks and training routines are implemented in TensorFlow 2 \cite{abadi2016tensorflow}.
For MLPs, we use the same number of neurons in each hidden layer.
For the convolutional networks, a single convolutional layer is followed by densely connected layers.
For all neural networks, we are using the ReLU activation function in the hidden layers. 
For the classifier we use the softmax activation in the output layer.
For the regression models, we use the hardSigmoid activation function in the output layer,

\begin{equation*}
    hardSigmoid(x) = 
    \begin{cases}
        0.0, \ & \text{if} \ \ x < -2.5, \\
        0.2x+0.5, \ & \text{if} \ \ -2.5 \leq x \leq 2.5, \\
        1.0, \  & \text{if} \ \ x > 2.5.
    \end{cases}
\end{equation*}

The network weights are initialized with the Xavier normal initialization \cite{Glorot2010}.
The training data is provided in Tab. \ref{datasets_table}.
We use the Adam optimizer \cite{Kingma2015} with a batch size of $1000$.
Regression models are trained for $200$ epochs.
We start with a learning rate with $1e-3$ and divide it by $10$ after $50, 100,$ and $150$ epochs. 
Classification models are trained for $10$ epochs with a constant learning rate of $1e-3$.




\section{Regression neural network for curvature}
\label{appendix_curvature}
In the Volume-of-Fluid method, the phases are distinguished by an abruptly-varying volume fraction field. This scalar field is not suitable for direct evaluation
of the curvature using discrete schemes. In the Levelset method on the other hand, the interface is tracked by a scalar field whose values represent 
the signed distance to the interface. The curvature can be directly evaluated from the smooth levelset field $\phi$ by discretizing the equation 
\eqref{curvature}, e.g. with central finite differences (FD). With this approach, the curvature is approximated at the cell center. Using equation \eqref{curvature_correction},
this value is corrected for the distance of the cell center from the interface by approximating the interface locally by a sphere. Here $d$ is the spatial dimension, thus in 
the present case $d=2$.

\begin{align}
    \kappa &= - \nabla\cdot \frac{\nabla \phi}{| \nabla \phi |} \label{curvature} \\
    \kappa_I &= \frac{(d - 1)\kappa}{d-1-\phi\kappa} \label{curvature_correction}
\end{align}

We train multilayer perceptrons to predict the curvature, similar to what is presented for the volume fraction and apertures (compare section \ref{methodology}).
From experience, we have seen that to achieve a predictive accuracy that is competitive to central finite differences, it is necessary to restrict the training data
to only concave (or convex) interfaces, i.e. only positive (or negative) curvature values.
In particular, we compare two training sets: The first one consists of only circular interfaces with $R_c=[2,100]$. The second one contains the first one plus
ellipsoidal interfaces with $a_e=[2,50]$ and an aspect ratio $[2,4]$. For interpretation of the paremeters and a detailed description of the sample generation, the reader is referred to the main text (Section \ref{data generation}).

\begin{figure}[!t]
    \centering
\begin{tikzpicture}
    \definecolor{color0}{rgb}{0.933333333333333,0.509803921568627,0.933333333333333}
    \definecolor{color0}{rgb}{0.933333333333333,0.509803921568627,0.933333333333333}
    \definecolor{color1}{rgb}{1,0.0784313725490196,0.576470588235294}
    \begin{groupplot}[group style={group size=3 by 1, horizontal sep = 0.5cm, vertical sep=1cm}, width=5cm, height=4cm, scale only axis]

        \nextgroupplot[
            legend cell align={left},
            legend style={draw opacity=1, text opacity=1, nodes={scale=0.7}, anchor=south west, at={(0.05,0.05)}},
            log basis x={10},
            log basis y={10},
            tick align=outside,
            tick pos=left,
            title={(a)},
            xmode=log,
            xtick style={color=black},
            ylabel={$MAE$},
            xlabel={$R$},
            ymin=1e-8, ymax=1,
            xmin=2.7, xmax=100,
            ymode=log,
            ytick style={color=black}
            ]
            \addplot [semithick, red, mark=*, mark size=1, mark options={solid}]
            table {%
            3.98107170553497 5.20135723001894e-05
            7.94328234724281 3.87816212518075e-05
            15.8489319246111 3.56142593997789e-05
            31.6227766016838 2.06611769790799e-05
            63.0957344480193 1.13912384009852e-05
            };
            \addplot [semithick, blue, mark=*, mark size=1, mark options={solid}]
            table {%
            3.98107170553497 0.00123577372527555
            7.94328234724281 0.000538738926636909
            15.8489319246111 0.000187623113995743
            31.6227766016838 0.000147838731463447
            63.0957344480193 0.000106670947863596
            };
            \addplot [semithick, black, mark=x, mark size=2, mark options={solid}]
            table {%
            3.98107170553497 0.00236586304399991
            7.94328234724281 0.000275302637533511
            15.8489319246111 3.56966978965074e-05
            31.6227766016838 4.47761543064273e-06
            63.0957344480193 5.56572913014103e-07
            };

        \nextgroupplot[
            legend cell align={left},
            legend style={draw opacity=1, text opacity=1, nodes={scale=0.85}, anchor=south west, at={(0.05,0.05)}},
            log basis x={10},
            log basis y={10},
            tick align=outside,
            tick pos=left,
            title={(b)},
            xlabel={$R$},
            xmin=2.72270130807791, xmax=11614.4861384034,
            xmode=log,
            ymin=1e-8, ymax=1,
            xmin=2.7, xmax=100,
            ymode=log,
            ytick style={color=black},
            yticklabels={,,}
        ]
        \addplot [semithick, red, mark=*, mark size=1, mark options={solid}]
        table {%
        3.98107170553497 0.160340634939739
        7.94328234724281 0.0341079946264405
        15.8489319246111 0.00497221746963137
        31.6227766016838 0.000640257474004682
        63.0957344480193 0.000354449897311783
        };
        \addlegendentry{\textit{c 3 100}}
        \addplot [semithick, blue, mark=*, mark size=1, mark options={solid}]
        table {%
        3.98107170553497 0.00390067147838751
        7.94328234724281 0.00111555978503278
        15.8489319246111 0.000470072264590265
        31.6227766016838 0.000292451873256941
        63.0957344480193 0.000190911504176227
        };
        \addlegendentry{\textit{ce 3 100}}
        \addplot [semithick, black, mark=x, mark size=2, mark options={solid}]
        table {%
        3.98107170553497 0.0843320006799435
        7.94328234724281 0.0181281816232977
        15.8489319246111 0.00300969178095892
        31.6227766016838 0.000392294411095641
        63.0957344480193 5.41802422431469e-05
        };
        \addlegendentry{FD}

    \end{groupplot}
    
    \end{tikzpicture}
    
    \caption{Mean absolute error $MAE$ of neural network and finite difference curvature to the exact solution over the interface resolution for a dataset consisting of
    (a) circular interfaces only and (b) circular and ellipsoidal interfaces. The red and blue line represent the neural networks trained using only circular and using circular 
    and ellipsoidal interface shapes, respectively. The black line indicates the central finite difference. For interpretation of the model architectures see section \ref{section:combined_model}.}
    \label{curvature_convergence}
\end{figure}
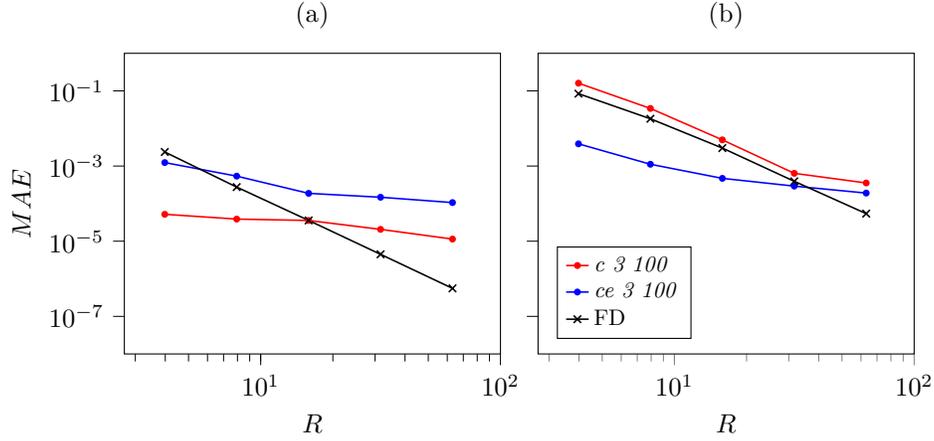

In Figure \ref{curvature_convergence}, the mean absolute error $MAE$ of the neural network and second order central finite difference to the exact curvature over the radius $R$
of the interface structures is displayed. Note that the radius $R$ is normalized using the cell size, i.e. the larger the radius the higher the resolution of the interface.
It can be seen, that the model trained using only circles achieves accuracy improvements until $R=15$ on circular interfaces. When including ellipses 
in the training set, the neural network accuracy is outperformed by the finite differences on circular interfaces practically over the whole resolution range. There is a tradeoff between 
accuracy on circular interfaces and generalization to different interface shapes. To make these models feasible within a solver, it is necessary to achieve 
improvements both on circles and on other interface shapes.

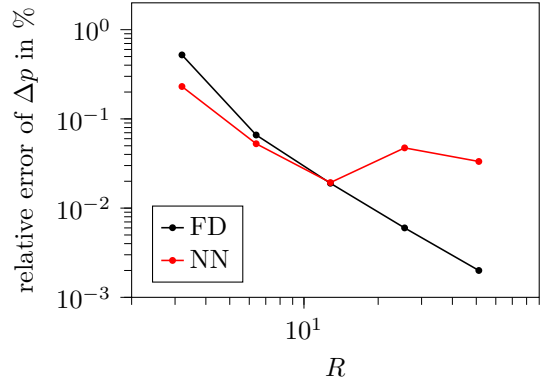
\begin{figure}[!t]
    \centering
\begin{tikzpicture}

    \definecolor{color0}{rgb}{0.12156862745098,0.466666666666667,0.705882352941177}
    \definecolor{color1}{rgb}{1,0.498039215686275,0.0549019607843137}
    \definecolor{color2}{rgb}{0.172549019607843,0.627450980392157,0.172549019607843}
    
    \begin{axis}[
        legend cell align={left},
        legend style={draw opacity=1, text opacity=1, nodes={scale=1.0}, anchor=south west, at={(0.05,0.05)}},
        log basis x={10},
        log basis y={10},
        tick align=outside,
        tick pos=left,
        width=7cm,
        height=5.5cm,
        xmode=log,
        xtick style={color=black},
        ylabel={relative error of $\Delta p$ in $\%$},
        xlabel={$R$},
        ymin=1e-3, ymax=2,
        xmin=2, xmax=0.9e2,
        ymode=log,
        xtick = {1e1},
        minor xtick={2,3,4,5,6,7,8,9,10,20,30,40,50,60,70,80,90},
        ytick style={color=black}
    ]
    \addplot [semithick, mark=*, mark size=1, black]
    table {%
    3.2 0.521333333333329
    6.4 0.0660000000000006
    12.8 0.0190000000000055
    25.6 0.00599999999999454
    51.2 0.0019999999999981
    };
    \addlegendentry{FD}
    \addplot [semithick, mark=*, mark size=1, red]
    table {%
    3.2 0.230666666666669
    6.4 0.0526666666666642
    12.8 0.0193333333333309
    25.6 0.0473333333333358
    51.2 0.0333333333333333
    };
    \addlegendentry{NN}
    \end{axis}
    
    \end{tikzpicture}
    
    \caption{Relative error of the pressure jump $\Delta p$ in the Young Laplace case for neural network and finite differences to the exact solution over the interface resolution.}
    \label{laplace}
\end{figure}

We implement the curvature neural network into a CFD solver and test the performance on a Young Laplace, i.e. a static drop with surface tension.
The exact pressure jump across the interface is given by $\Delta p = \sfrac{\sigma}{R_0}$, where $\sigma$ is the surface tension coefficient and $R_0$ is 
the drop radius. The relative error for neural network and finite differences over the interface resolution is shown in Figure \ref{laplace}. In particular, we
are using the neural network that has been trained only on circular interfaces $\textit{c 3 100}$. To compute the pressure jump in the CFD, the mean pressure within the 
drop is used. In accordance to the convergence tests discussed previously, accuracy improvements until $R\approx12$ are observed. 

\newpage

\bibliography{bib}

\end{document}